\documentclass[paper=a4, fontsize=12pt]{aastex631}

\usepackage{amsmath}
\usepackage{xcolor}
\graphicspath{{./}{figures/}}
\begin{document}

\title{Entropy Stabilized ZrHfCoNiSnSb Half-Heusler Alloy for Thermoelectric Applications: A Theoretical Prediction}


\author{Rajeev Ranjan}
\affiliation{Department of Physics, Indian Institute of Science Education and Research, Pune 411008, India}
\email{rajeev.ranjan@students.iiserpune.ac.in} 
\begin{abstract}

Half-Heusler (HH) alloys are potential thermoelectric materials for use at elevated temperatures due to their high Seebeck coefficient and superior mechanical and thermal stability. However, their enhanced lattice thermal conductivity is detrimental to thermoelectric applications. One way to circumvent this problem is to introduce mass disorder at lattice sites by mixing the components of two or more alloys. Such systems are typically stabilized by the entropy of mixing.
In this work, using computational tools, we propose a mixed HH, namely, ZrHfCoNiSnSb, which can be formed by the elemental compositions of the parent half-Heuslers ZrNiSn/HfNiSn and HfCoSb/ZrCoSb. We propose that this new compound can be synthesized at elevated temperatures, as its Gibbs free energy is reduced due to higher configurational entropy, making it more thermodynamically stable than the parent compounds under such conditions.
Our calculations indicate that it is a dynamically stable semiconductor with a band gap of 0.61 eV. Its lattice thermal conductivity at room temperature is $5.39~\text{Wm}^{-1}\text{K}^{-1}$, which is significantly lower than those of the parent compounds. The peak value of this alloy's figure of merit (ZT) is 1.00 for the n-type carriers at 1100 K, which is 27\% more than the best figure of merit obtained for the parent compounds.

\end{abstract}

 \section{Introduction}
 An enormous amount of waste heat gets generated in the automotive exhaust, home heating, and industrial processes. One of the ways of utilizing this wasted heat is to convert it to electricity by using the phenomena of thermoelectricity.
 The efficiency of a thermoelectric material is determined by its figure of merit(ZT), which is a function of Seebeck coefficient ($\alpha$), electrical conductivity ($\sigma$), thermal conductivity ($k_{t}$), and the temperature (T). This figure of merit is given by the relation \cite{snyder2008complex} :

 \begin{equation}
 \text{ZT} = \frac{\alpha{^2}\sigma}{k_{t}}\text{T}
   \label{eq:eq_zt}
 \end{equation}
 
 Thermal conductivity ($k_{t}$) has contributions from both electrons ($k_{e}$) and phonons ($k_{L}$). To achieve a high value of ZT, which is required for being a good thermoelectric material,  a high Seebeck coefficient, a high electrical conductivity, and a low thermal conductivity are required.
 
\begin{figure}[h!]

 		  		 	\centering
 		  		 	\begin{minipage}{0.70\linewidth}

 		  		 		\includegraphics[width=0.90\textwidth]{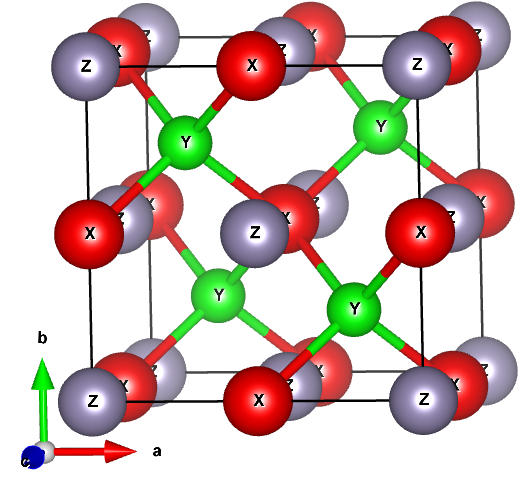}
 		  		 		
 		  		 		\label{fig:minipage1}
 		  		 	\end{minipage}
 		  		 	
 		  		 	\caption{Crystal structure of half-Heusler alloy XYZ}
 		  		\label{fig:structure_xyz}
 		  		 \end{figure}

 Half-Heusler alloys are one class of intermetallic compounds that exhibit great promise as thermoelectric materials suitable for high-temperature applications owing to their remarkable attributes, including a high Seebeck coefficient, exceptional mechanical strength, and thermal stability. On the contrary, the increased lattice thermal conductivity of these materials poses a disadvantage for thermoelectric applications. These compounds, having a composition of XYZ  where X (Wyckoff position 4b(0.5, 0.5, 0.5)) and Y (Wyckoff position 4c(0.25, 0.25, 0.25)) are transition metals and Z (Wyckoff position 4a(0, 0, 0)) is a p-block element, comprise of three interlocking face-centred cubic sublattices and an additional vacant sublattice in the same cubic structure. The semiconducting properties and stability of half-Heusler compounds can be understood using the Zintl concept \cite{zeier2016engineering}, according to which the most electropositive element, X, acts as a cation and donates all of its valence electrons to the tetrahedrally bonded YZ sublattice, effectively forming the anionic part of the structure. Based on this concept, a half-Heusler compound with a valence electron count (VEC) of 18 may be a stable semiconductor with potential for thermoelectric applications. TiNiSn \cite{gandi2016electron,muta2009high}, ZrNiSn \cite{gandi2016electron,muta2009high}, HfNiSn \cite{gandi2016electron,zou2013electronic}, TiCoSb \cite{gandi2017thermoelectric,sekimoto2005thermoelectric}, ZrCoSb \cite{gandi2017thermoelectric,sekimoto2005thermoelectric} and HfCoSb \cite{gandi2017thermoelectric, sekimoto2005thermoelectric} are some of the well-studied VEC 18 half-Heuslers for thermoelectric applications. The crystal structure of half-Heusler alloy XYZ is shown in Figure~\ref{fig:structure_xyz}  \\

 		  		 \bigskip
 		  		 
 		  		  \ \ \ \ \ \
 
To overcome the issue of lattice thermal conductivity, a potential solution involves inducing mass disorder at lattice sites by blending the elemental compositions of two or more half-Heusler (HH) alloys, thereby forming a high-entropy alloy. The concept of high entropy has been applied to various classes of materials, including chalcogenides \cite{liu2017entropy, zhang2018data} and oxides \cite{jiang2018new,rost2015entropy}. In these systems, a high configurational entropy favors the formation of a single-phase structure by contributing significantly to the Gibbs free energy \cite{zhang2019review}. Compared to conventional solid solutions with low levels of elemental additions, the high-entropy effect can overcome limitations in solubility. This was demonstrated by the successful synthesis of a single-phase (MgCoNiCuZn)O compound with a rock-salt structure \cite{rost2015entropy}, which can be viewed as an equimolar mixture of MgO, CoO, NiO, CuO, and ZnO. In this system, the typical solubility limits of binary combinations such as MgO–ZnO and CuO–NiO were surpassed due to entropic stabilization. Among half-Heusler compounds, the high-entropy concept has also been successfully applied in the synthesis of a single-phase (TiZrHfVNbTa)Fe$_{0.5}$Co$_{0.5}$Sb \cite{chen2022synthesis}, which is based on six VEC 18 half-Heuslers: TiCoSb, ZrCoSb, HfCoSb, VFeSb, NbFeSb, and TaFeSb. Another notable example, though not a high-entropy alloy, is the synthesis of $\text{Ti}_{2}\text{NiCoSnSb}$ \cite{karati2019ti2nicosnsb}, which can be regarded as an equimolar mixture of the parent HHs TiNiSn and TiCoSb. This alloy has a lattice thermal conductivity of 7 Wm$^{-1}$K$^{-1}$ as compared to 24 Wm$^{-1}$K$^{-1}$ for TiCoSb \cite{sekimoto2005thermoelectric} and 13 Wm$^{-1}$K$^{-1}$ for TiNiSn \cite{gandi2016electron} at room temperature. However, its power factor is significantly reduced as compared to parent compounds, leading to the low value of the figure of merit. Intriguingly, the question arises as to whether it is feasible to synthesize a material by combining the compositions of two half-Heusler compounds with reduced lattice thermal conductivity while retaining the power factor of the original parent compounds. The investigation also seeks to ascertain the stability of this mixed compound in comparison to its parent compounds. Additionally, it will be interesting to understand how the blending process influences the electronic transport properties of this composite material.\\
 ZrHfCoNiSnSb, which can be seen as a mixture of ZrNiSn/HfNiSn and HfCoSb/ZrNiSn, is dynamically stable and predicted to be synthesized at high temperatures. The lattice thermal conductivity of this compound is found to be 5.39 Wm$^{-1}$K$^{-1}$ at 300 K, which is much lower than the parent compounds. Additionally, for the n-type case, its power factor lies midway between the power factors of the constituent parent compounds, resulting in a ZT value of 1.00 at a temperature of 1100 K.

\section{Computational details}
\label{sec:comp_det}

\begin{figure}[h]
    \centering
    \includegraphics[scale=0.35]{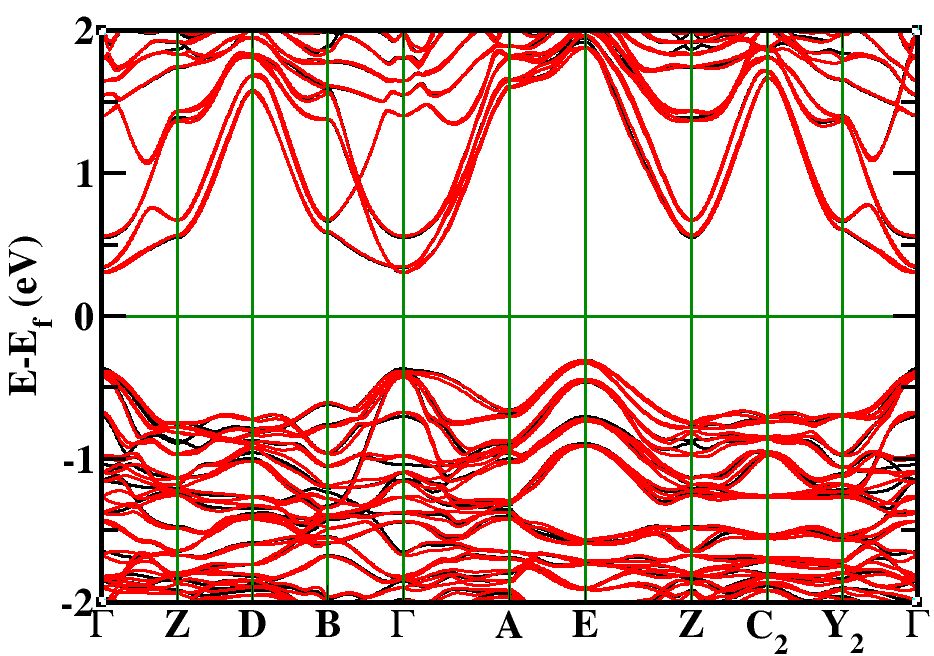}
    \caption{Band structure comparison of the HEA, showing  calculations without spin-orbit coupling (black) and with spin-orbit coupling (red)}
    \label{fig:bs_comparison}
\end{figure}

The calculations were carried out using plane-wave density functional 
theory (DFT) based calculations as implemented in the Quantum ESPRESSO \cite{giannozzi2009quantum, Giannozzi2017}
software. The electron-ion interactions were described using ultrasoft pseudopotentials. For the wavefunction (charge density), we have used a
basis set whose size corresponds to a kinetic energy cutoff of 60(480) Ry.
The electron-electron exchange and correlation effects were treated
using the Perdew-Burke-Ernzerhof (PBE) \cite{perdew1996generalized} 
parametrization of the generalized gradient approximation (GGA).
For electronic calculations, the Brillouin zone was sampled with a shifted $10\times10\times10$ 
and $6\times6\times6$ Monkhorst-pack $k$-mesh for the conventional unit cell of the parent compounds and ZrHfCoNiSnSb, respectively. To compute the density of states (DOS) 
and the electronic transport properties we have used 
$20\times20\times20$ $k$-mesh grid for the parent compounds and $18\times18\times18$ $k$-mesh grid for ZrHfCoNiSnSb. Since spin-orbit interaction has a negligible effect in the HEA (as evident from the band structure shown in Figure~\ref{fig:bs_comparison} of the SI), it was not included in the DFT calculations.

To study the dynamical stability, vibrational properties and
lattice thermal conductivity, the phonons were computed using density 
functional perturbation theory\cite{baroni2001phonons}. The calculations were 
performed on a $6\times6\times6$ $q$-mesh for the primitive unit cell of the parent compounds and a 
$3\times3\times3$ q-mesh for ZrHfCoNiSnSb.

Electronic transport properties were calculated by using the semi-classical 
Boltzmann transport theory within the constant relaxation time and rigid band 
approximations as implemented in the BoltzTrap2 code 
\cite{madsen2018boltztrap2}. Under these approximations the $(ij)^{th}$ component of projected conductivity
tensor per unit relaxation time $ \left( \frac{\sigma_{ij}}{\tau}  \right)$ 
was calculated as 
	 
\begin{equation}
  \frac{\sigma_{ij}(\epsilon)}{\tau}= e^{2} \sum_{\beta}\int\frac{d^{3}\vec{k}}{4\pi^{3}} \delta(\epsilon - \epsilon(\beta;\vec{k})) v_{i}(\beta;\vec{k}) v_{j}(\beta;\vec{k})
  \label{eq:proj_s}
\end{equation}

\noindent Here $e$ is the charge of the electron and $\tau$ is the constant 
relaxation time. $\vec{v}(\beta;\vec{k})=\frac{1}{\hbar}\nabla_{k}
(\epsilon(\beta;\vec{k}))$ is the group velocity of the electron
occupying the $\beta^{th}$ band at the $k^{th}$ k-point of the BZ and  
$\epsilon(\beta;\vec{k})$ is the energy eigenvalue corresponding to that electronic state. The $(ij)^{th}$ component of 
electrical conductivity per unit relaxation time $\left(\frac{\sigma_{ij}(T;\mu)}{\tau}\right)$, Seebeck coefficient $\left( \alpha_{ij}(T;\mu)\right) $ and electronic thermal conductivity per unit relaxation time $\left(\frac{\kappa_{ij}^{e} (T;\mu)}{\tau}\right) $ were calculated from
Equation~\ref{eq:proj_s} as: 
	 
\begin{equation}
  \frac{\sigma_{ij}(T;\mu)}{\tau} = \frac{1}{\Omega}\int d\epsilon
	 \left(-\frac{\partial f_{0}(T;\mu)}{\partial\epsilon}\right) \left(\frac{\sigma_{ij}(\epsilon)}{\tau}\right)
       \label{eq:sig_t_mu}
\end{equation}
	 
\begin{equation}
	 \alpha_{ij}(T;\mu) = \left(\frac{1}{eT\Omega}\right) \sum_{k}\left(\tau \sigma^{-1}_{ik}(T;\mu) \right) \left(\int d\epsilon \left(-\frac{\partial f_{0}(T;\mu)}{\partial\epsilon}\right)(\epsilon - \mu) \left(\frac{\sigma_{kj}(\epsilon)}{\tau}\right)\right)
       \label{eq:seebeck}
\end{equation}

\noindent and

\begin{equation}
	 \frac{\kappa_{ij}^{e} (T;\mu)}{\tau} = \frac{\kappa_{ij} (T;\mu)}{\tau} -T\sum_{\alpha, \beta}\nu_{i\alpha}\left(\frac{\sigma^{-1}_{\beta\alpha}}{\tau}\right)\nu_{\beta j}
       \label{eq:electron_kappa}
\end{equation}

\noindent where

\begin{subequations}
    \begin{equation}
	 \begin{aligned}
	  \frac{\kappa_{ij} (T;\mu)}{\tau} = \left(\frac{1}{e^{2}T\Omega}\right)\int d\epsilon \left(-\frac{\partial f_{0}(T;\mu)}{\partial\epsilon}\right) (\epsilon - \mu)^{2} \left(\frac{\sigma_{ij}(\epsilon)}{\tau}\right) 
	   \end{aligned}
         \label{eq:kapaa_e_part1}
     \end{equation}
      \noindent\text{and}
      \begin{equation}
	   \begin{aligned}
    \nu_{ij} =  \frac{1}{eT\Omega}\int d\epsilon \left(-\frac{\partial f_{0}(T;\mu)}{\partial\epsilon}\right)(\epsilon - \mu) \left(\frac{\sigma_{kj}(\epsilon)}{\tau}\right)
	  \end{aligned}
        \label{eq:kappa_e_part2}
      \end{equation}
\end{subequations}
	 
\noindent In these equations $f_{0}(T;\mu)$ is the equilibrium Fermi-Dirac distribution at 
temperature $T$ and chemical potential $\mu$ and $\Omega$ is the unit cell
volume. The values of $\tau$ for electrons and holes,
appearing in the above equations, were computed using deformation potential theory\cite{bardeen1950deformation} as described in Section~\ref{sec:eff_mass} and \ref{sec:tau}.
	 
\section{Results and discussions}
\label{sec:res_dis}

\subsection{Structure, thermodynamic stability and bonding}
\label{sec:str_ther_bond}	  

\begin{figure}[h!]
    \centering
\includegraphics[width=\columnwidth]{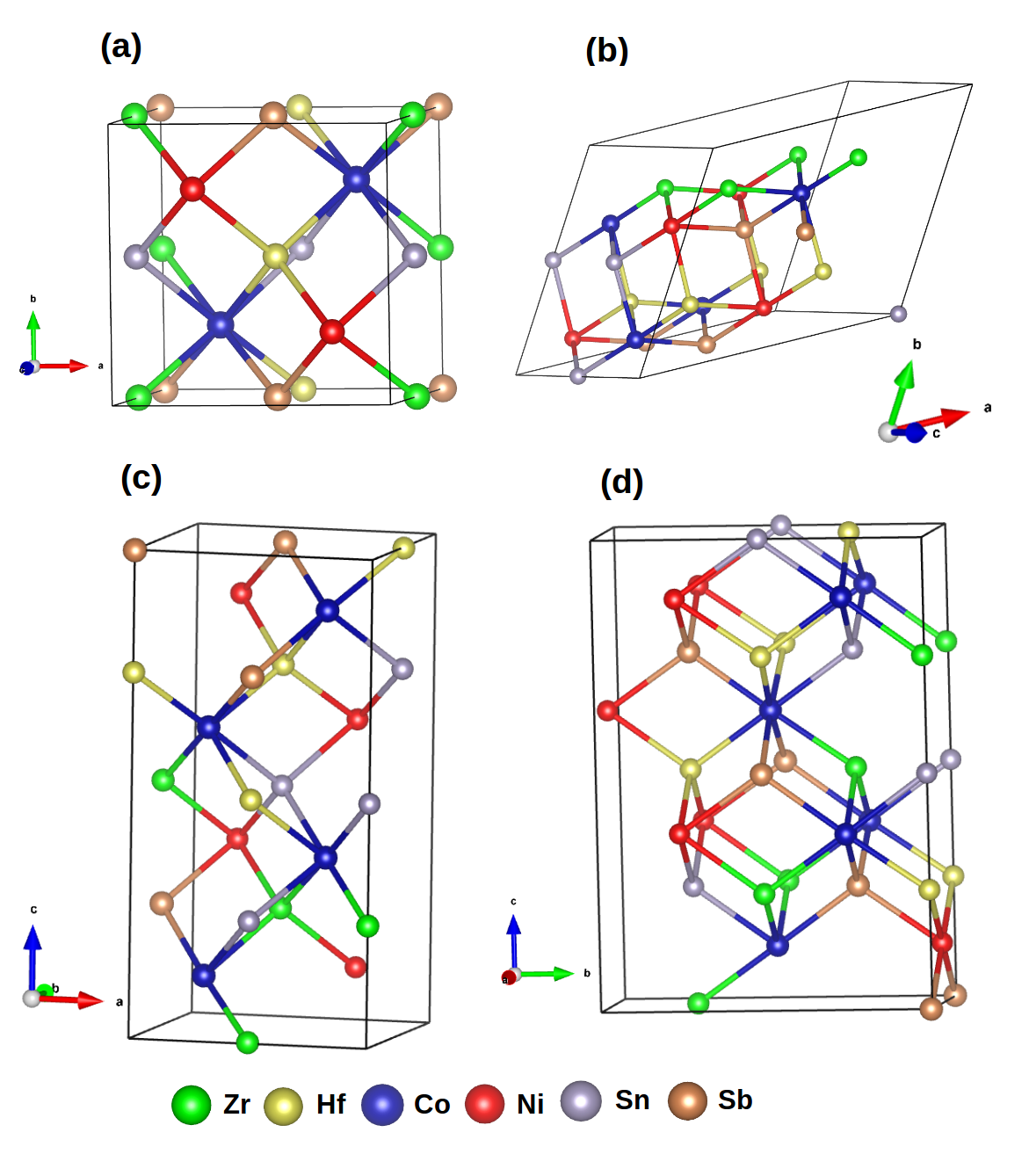} 
 	\caption{Crystal structure of the optimized SQSs (a) Conventional unit cell (b) $2\times 2 \times 2$ supercell  (c) Double conventional unit cell (d) unconstrained SQS with 24 atoms}
    \label{fig:structure_po}
\end{figure}

\begin{figure}[h]
    \centering
\includegraphics[width=\columnwidth]{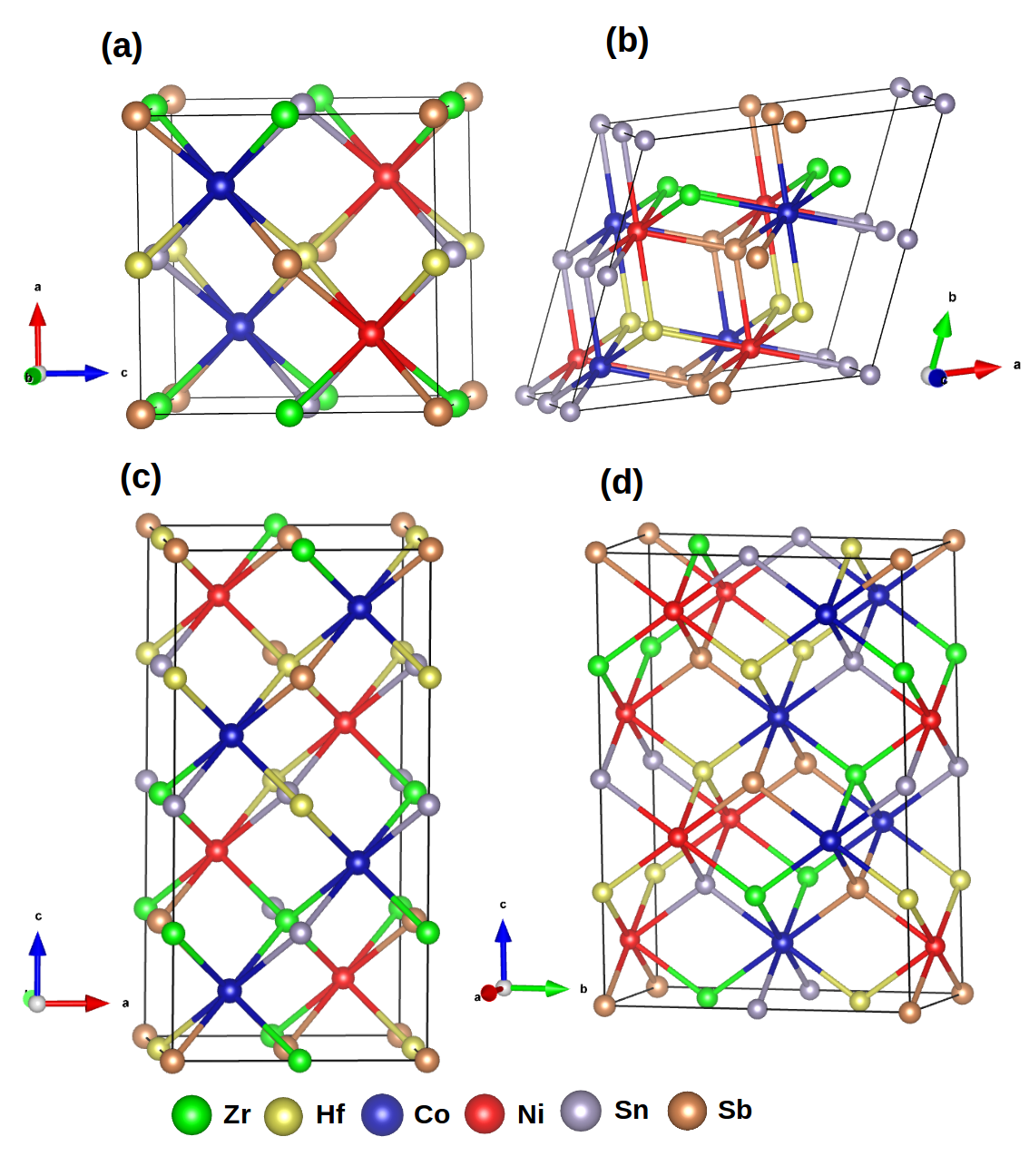} 
	\caption{Crystal structure of the generated SQSs before optimization (a) Conventional unit cell (b) $2\times 2 \times 2$ supercell  (c) Double conventional unit cell (d) unconstrained SQS with 24 atoms}
    \label{fig:strucctures_bo}
\end{figure}

To model the most disordered configuration of ZrHfCoNiSnSb, we employed the Monte Carlo Special Quasirandom Structure (McSQS) method, as implemented in the ATAT software\cite{van2013efficient}. The quasirandom
structures were generated with the constraint that the Zr and Hf atoms
occupy the Wyckoff site 4b (0.5, 0.5, 0.5), Ni and Co atoms occupy
the site 4c (0.25, 0.25, 0.25) and Sn and
Sb atoms occupy the site 4a (0, 0, 0). We constructed four different quasirandom structures within this constraint. Three of these structures are the 12-atom
structure in the conventional cubic HH supercell (Type I), the 24-atom quasirandom structure with the conventional cubic cell doubled along the c-
direction (Type II), and the 24-atom quasirandom structure with a 
2$\times$2$\times$2 primitive fcc supercell (Type III). 
Additionally, we also generated a 24-atom SQS with no constraints on the particular choice of crystal structure (Type IV).
The initial structures of these unit cells are shown in Figure~\ref{fig:strucctures_bo} of the
 Supporting Information (SI), while the optimized structures are shown in Figure~\ref{fig:structure_po}

On optimizing the lattice parameters and the atomic 
positions of the above-mentioned unit cells, we find that the Type I structure 
is the lowest in energy. The relative energies per formula unit of the other 
structures with respect to that of Type I ($\Delta E$) and the lattice 
parameters of all the structures are reported in 
Table~\ref{table:opt_str_en}. It is interesting to note that post-geometry
optimization, the lattice parameters and the angles between the lattice
vectors deviate from those observed in conventional cubic of fcc lattice.
 For example, in the lowest-energy Type I structure, all three lattice vectors are equal, and the interaxial angles are 90.000° prior to geometry optimization. After optimization, the structure relaxes into a monoclinic phase characterized by the $C_{2h} (2/m)$ point group, due to slight variations in the lattice parameters and a deviation of the angle $\gamma$ from 90.000$^{\circ}$. Since Type I is the
lowest energy structure amongst the ones considered in this study, all 
subsequent calculations for ZrHfCoNiSnSb  were performed using this
unit cell.
	  
	  
	  
	  
	  
\begin{table} [ht]
    \centering 
     \caption{Computed relative energy per formula unit
     ($\Delta E$), lattice parameters and angle between the lattice parameters for the different SQS structures considered in this
     study.}
     \begin{tabular}{|c|c|c|c|c|}
	\hline
	  Structure & Type-I & Type-II & Type-III & Type-IV \\
	  \hline 
	\shortstack{$\Delta E$ (in eV)} & 0.000 & 0.028 & 0.014 & 0.026  \\
	  \hline
       \shortstack{Lattice \\ parameters  \\  (in $\AA$)}  & \shortstack{a \ = \ 6.109, \\  b \ = 6.107, \\ \ c \ = 6.115 \ }  & \shortstack{a \ = \ 6.101, \\ b \ = 6.108, \\ \ c \ = 12.199 \ }  & \shortstack{a \ = \ 8.623, \\ b \ = 8.628, \\ \ c \ = 8.626 \ }  & \shortstack{a \ = \ 4.318, \\ b \ = 8.626, \\ \ c \ = 12.204 \ } \\
	  		\hline
      Angles  & \shortstack{$\alpha$ \ = \ $\beta$ \ = $90.000^{\circ}$,  \\ $\gamma$ \ =\  $90.004^{\circ}$} 
	  		 & \shortstack{$\alpha$ \ =  $90.006^{\circ}$, \\ $\beta$ \ = $90.001^{\circ}$,  \\ $\gamma$ \ =\  $89.979^{\circ}$}
	  		  & \shortstack{$\alpha$ \ = $60.084^{\circ}$ \\ $\beta$ \ = $60.065^{\circ}$,  \\ $\gamma$ \ =\  $60.075^{\circ}$}
	  		   & \shortstack{$\alpha$ \ = $90.044^{\circ}$ \\ $\beta$ \ = $90.000^{\circ}$,  \\ $\gamma$ \ =\  $90.000^{\circ}$}  \\
	  		\hline		
    \end{tabular}
    \label{table:opt_str_en}
\end{table}

These alloys are typically synthesized experimentally via arc melting of the elemental precursors. Hence to check the stability of the HEA against
 segregation into the individual bulk components, we have computed their
 average formation energy ($E_f$) which is given as:
 
 \begin{equation}
     E_f=\frac{E_{\text{ZrHfCoNiSnSb }}-aE_{\text{Hf}}-bE_{\text{Zr}}-cE_{\text{Ni}}-xE_{\text{Co}}-yE_{\text{Sn}}-zE_{\text{Sb}}}{(a+b+c+x+y+z)}
    \label{eq:eqn_ef}
 \end{equation}
 
 \noindent where $E_{\text{ZrHfCoNiSnSb }}$ is the total energy of the HEA and
 $E_i$, $i=$Hf, Zr, Ni, Co, Sn and Sb, are the energy per atom of the
 $i^{th}$ element in its bulk. $a$, $b$, $c$, $x$, $y$ and $z$ are the number
 of atoms of Hf, Zr, Ni, Co, Sn and Sb, respectively, in the HEA. We find
 that our proposed HEA has a formation energy of -4.48 eV per formula unit, suggesting
 that this is highly stable with respect to the segregation into individual
 atomic phases.

 Additionally, the proposed HEA might also be thought of as a mixture of
 two stable HHs, namely, HfNiSn and ZrCoSb HHs or ZrNiSn and HfCoSb HHs. Hence, it is also imperative
 to study the stability of the HEA with respect to phase segregation
 into these HHs. To do so, we considered the following chemical reactions:

\begin{equation}
    \text{ZrHfCoNiSnSb } \longrightarrow \text{HfNiSn} + \text{ZrCoSb} 
\label{rect1}
\end{equation}
\begin{equation}{1}
\text{ZrHfCoNiSnSb }  \longrightarrow \text{ZrNiSn} + \text{HfCoSb}
\label{rect2}
\end{equation}

\noindent and computed the enthalpy of formation ($\Delta H_{e}$), which is
given by:
	
\begin{equation}
\Delta H_{e} = E_{\text{HfNiSn/ZrNiSn}} + E_{\text{ZrCoSb/HfCoSb}}
-\ E_{\text{ZrHfCoNiSnSb }}
\label{eq:delta_H}
\end{equation}

\noindent where the first, second and third terms on the right-hand side
of Eqn.~\ref{eq:delta_H} are the total energies of the HHs into which 
they can phase segregate and HEA, respectively. For 
reactions~\ref{rect1} and ~\ref{rect2}, we obtain $\Delta H_{e}$ to be -70 meV and -86 meV per formula unit. While the negative 
values of $\Delta H_{e}$ might initially suggest that ZrHfCoNiSnSb  will phase
segregate to either HfNiSn and ZrCoSb or ZrniSn and HfCoSb, the role of 
configurational entropy ($\Delta S_{\text{config}}$) at elevated synthesis 
temperatures become crucial in reducing the Gibbs free energy  and 
contributing to the overall thermodynamic stability \cite{zhang2019review} of ZrHfCoNiSnSb  over 
HfNiSn/ZrNiSn and ZrCoSb/HfCoSb. This configurational entropy is given by\cite{chen2022synthesis}:
	  	  
\begin{equation}
\Delta S_{\text{config}} \ = -k_{B} \left(\sum_{x=1}^{m}\sum_{i=1}^{n}f_{i}^{x}ln\left(f_{i}^{x}\right)\right)
\label{eq: entropy}
\end{equation} \\

\noindent In Eqn~\ref{eq: entropy}, $k_{B}$ is the Boltzmann constant. The 
summation over $x$ runs over all the sublattices (here it is 3) and 
$f_{i}^{x}$ is the fraction of element $i$ in the sublattice $x$. For the 
case of ZrHfCoNiSnSb, in Type-I, each of  Hf/Zr, Ni/Co, and Sn/Sb forms fcc 
sublattice. Using Eqn.~\ref{eq: entropy}$, \Delta S_{\text{config}}$ for the Type-I
structure of the HEA comes out to be 2.079  $k_{b}$. The temperature above 
which this system can be synthesized will be the one at which the change
in the Gibbs free energy ($\Delta G$) is positive. 
This Gibbs free energy difference, $\Delta G$, is expressed as:
\begin{equation}
    \Delta G = \Delta G_{e} + \Delta G_{p}
    \label{eq:den_g_full}
\end{equation}
where the contributions of electronic energy and configurational entropy to the Gibbs free energy, $\Delta G_{e}$, is given by
\begin{equation}
    \Delta G_{e} = \Delta H_{e} - T\Delta S_{\text{config}}
    \label{eq:delg_e}
\end{equation}
and $\Delta G_{p}$ represents the phonon contribution to the Gibbs free energy, defined as:
\begin{equation}
    \Delta G_{p} = \Delta H_{p} - T\Delta S_{p}
    \label{eq:delg_p}
\end{equation}

Here, $H_{p}$ and $S_{p}$ denote the phonon energy and phonon entropy of the respective structures, given by:
\begin{equation}
    H_{p} = \int_0^{\omega_{\max}} g(\omega) \hbar \omega \left( \frac{1}{e^{\hbar \omega / k_B T} - 1} + \frac{1}{2} \right) d\omega
 \label{eq:eqn_hp}
\end{equation}

and
\begin{equation}
    S_{p} = \int_0^{\omega_{\max}} g(\omega) k_B \left[ \frac{\hbar \omega / k_B T}{e^{\hbar \omega / k_B T} - 1} - \ln\left( 1 - e^{-\hbar \omega / k_B T} \right) \right] d\omega       
    \label{eq:eqn_sp}
\end{equation}

In these expressions, $g(\omega)$ represents the phonon density of states corresponding to the frequency $\omega$, and $\omega_{\max}$ is the maximum phonon frequency.

\begin{figure}[h!]
    \centering
    \begin{minipage}{0.95\linewidth}
        \includegraphics[width=0.90\textwidth]{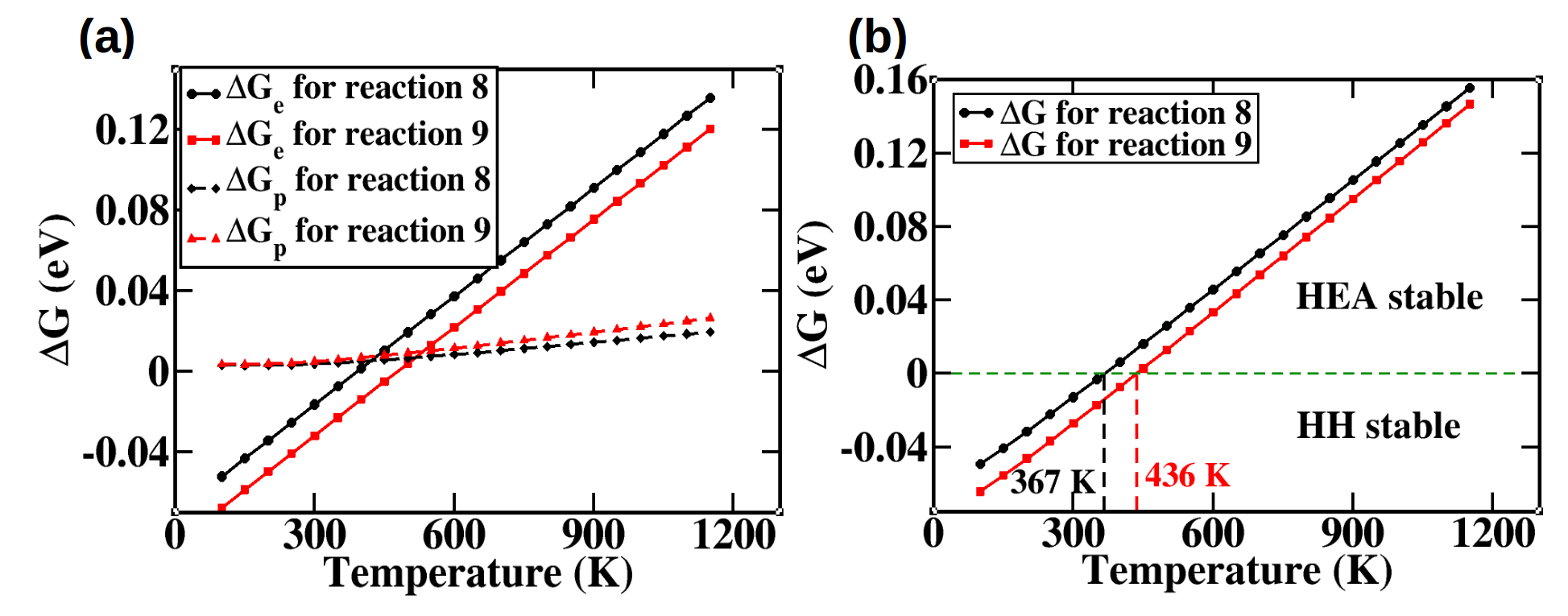}
        \label{fig:minipage1}
    \end{minipage}
    \caption{(a) Individual components of $\Delta G$ as a function of temperature. (b) Total Gibbs free energy as a function of temperature.}
    \label{fig:delta_g}
\end{figure}

Figure~\ref{fig:delta_g} illustrates the variation of $\Delta G$ with temperature for the cases described in Equations~\ref{rect1} and \ref{rect2}.

\noindent As can be seen from the Figure~\ref{fig:delta_g}, $\Delta G$ is positive when $T$ is greater than 367 K and 436 K for reactions~\ref{rect1} and
\ref{rect2}, respectively. We note that these temperatures are only 67 K
and 136 K above room temperature and are much lower than the
synthesis temperatures of half-Heusler systems. Thus, our analysis 
suggests that these materials will be stable towards segregation into the HHs.

At this point, it would be interesting to compare the local geometry of this
HEA with those of conventional HHs. A conventional HH
can be thought of as a combination of two zinc blend structures formed by the YZ elements and the XY elements. In each of the zinc blende
structures the Y element is at the centre of the tetrahedra
formed either by the X or the Z element. Moreover, these are perfect tetrahedra, i.e., in each tetrahedra, the four Y-Z or Y-X bonds
have the same bond lengths. In contrast, for the HEA, the introduction of the 
disorder at each of the atomic sites distorts these tetrahedra because 
two of the vertices are occupied by Hf (Sb) while the other two by Zr 
(Sn). In ZrNiSn and HfNiSn the Ni-Sn bond lengths are 2.665 \AA~and
2.647 \AA, respectively while in the HEA we find the bond length to
be 2.656 \AA. Similarly, the Co-Sb bond length in HEA is 2.625 \AA~which
is slightly shorter (similar) than (to) that of 2.642 \AA~(2.624 \AA)
in ZrCoSb (HfCoSb). Moreover, we also observe formation of new
Ni-Sb and Co-Sn bonds with bond lengths of 2.625 \AA~and 2.662 \AA~,
respectively. In the X-Y sublattice of the HEA the Zr-Ni (Zr-Co) bonds
are elongated (shortened) compared to that observed in ZrNiSn and
ZrCoSb ($d_{Zr-Ni}$=2.688 \AA~in HEA vs 2.665 \AA~in ZrNiSn;
$d_{Zr-Co}$=2.609 \AA~in HEA vs. 2.642 \AA~in ZrCoSb).
In contrast, both the Hf-Ni and Hf-Co bonds in HEA are shortened
compared to that observed in the parent HHs. However, the Zr-Ni
(Zr-Co) bond is elongated (shortened) than that observed in ZrNiSn
(ZrCoSb). These asymmetries in the local
geometry result in deviation from the cubic symmetry. Further, such a 
wide variation of bond lengths suggests that there is significant 
heterogeneity in terms of bonding and bond strength in the HEA, the 
implications of which on the thermoelectric properties are discussed 
later.

\bigskip

\begin{figure}
    \centering
    \includegraphics[scale=0.35]{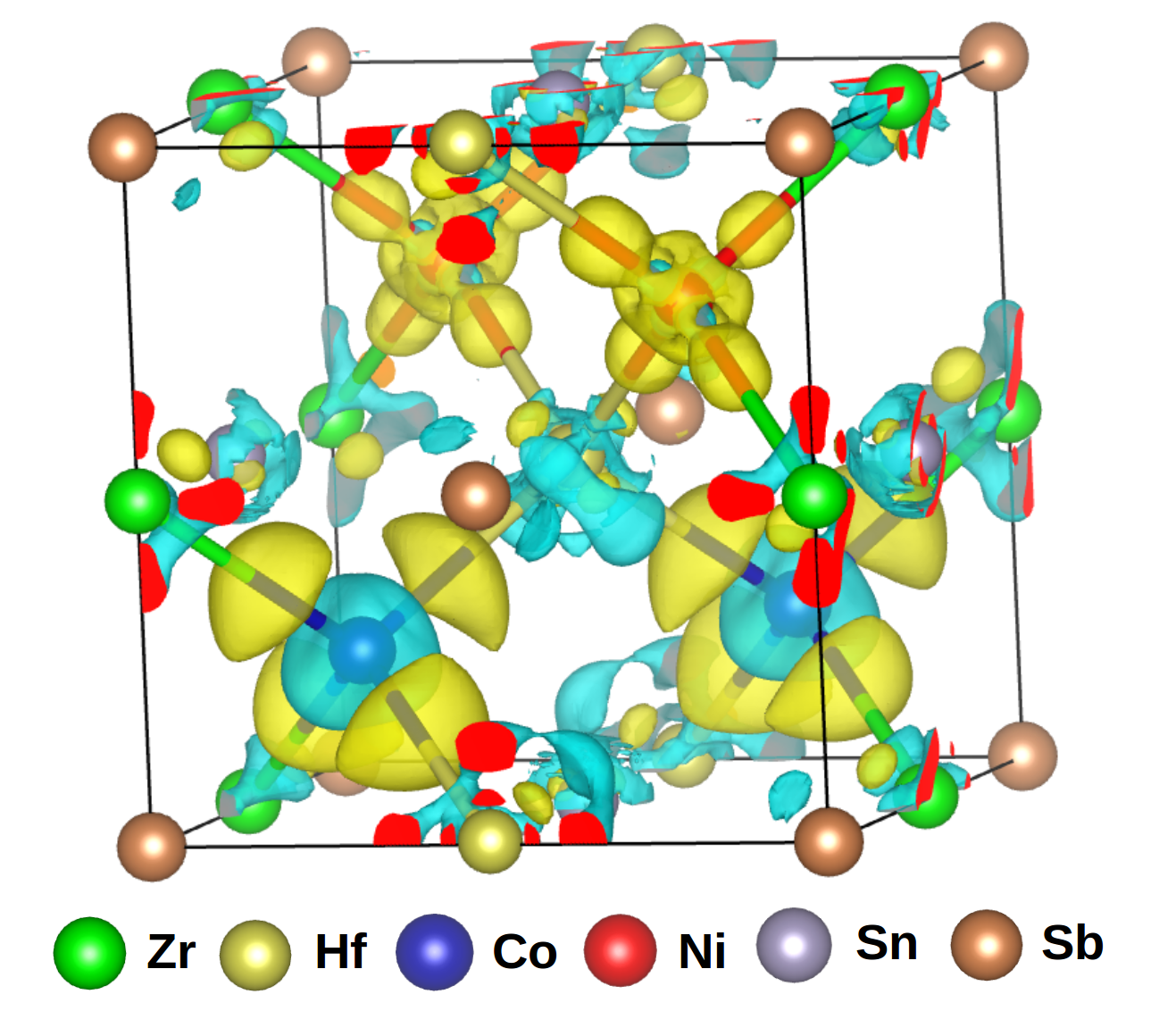}
    \caption{Isosurfaces of charge density difference ($\Delta \rho$) between the
    electron density of the HEA and that obtained from the superposition of the atomic densities. Yellow (Turquoise) isosurfaces denote accumulation (depletion) of charge density. }
    \label{fig:delta_rho_hea}
\end{figure}

\begin{figure}[h]
    \centering
\includegraphics[width=\columnwidth]{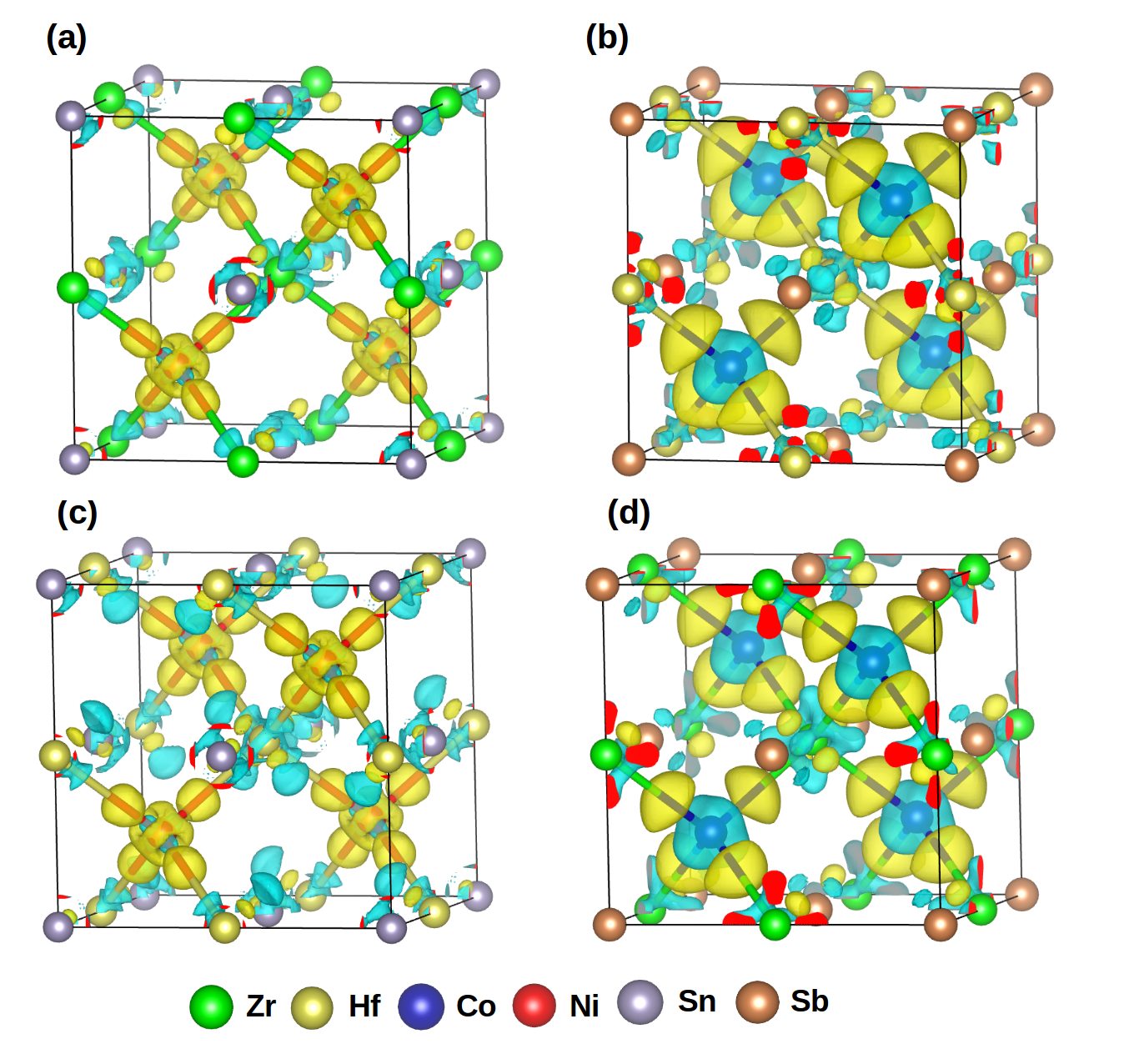} 

    \caption{Isosurfaces of charge density difference ($\Delta \rho$) between the
    electron density of the parent compounds and that obtained from the superposition of the atomic densities. Yellow (Turquoise) isosurfaces denote accumulation (depletion) of charge density (a) ZrNiSn (b) HfCoSb (c) HfNiSn (d) ZrCoSb }
    
  \label{fig:del_rho_hh}
\end{figure}

In order to understand the nature of the bonding between the different
elements in the HEA, we have computed the difference between the charge
density distribution and superposition of the atomic 
densities ($\Delta \rho$) for the HEA and the parent compounds.
$\Delta \rho$ provides information as to how the atomic charge 
densities are rearranged when the different elements interact to form a 
compound. Figure~\ref{fig:delta_rho_hea} shows the $\Delta \rho$ for 
the HEA while those for the parent compounds are shown in Figure~\ref{fig:del_rho_hh}  of 
the SI. In the Co-containing parent compounds, namely ZrCoSb and 
HfCoSb, we observe that there is charge depletion from the Co atom 
(Turquoise isosurfaces in Figure~\ref{fig:del_rho_hh} (b) and (d) of SI). Further, both charge depletion and 
accumulation can be observed around the Hf/Zr atoms, with the former 
dominating. Charge depletion is also observed around Sb. 
Importantly, we observe charge accumulation in between the Zr/Hf and Co 
bonds suggesting a covalent nature of bonding between them. No such 
charge accumulation is observed between the Co-Sb bonds. Similarly, in 
the Ni-containing compounds, i.e., in ZrNiSn and HfNiSn, we primarily 
observe charge depletion around Zr/Hf and Sn. However, in contrast with Co, we 
observe both accumulation and depletion of charges from Ni, with the 
former dominating. Additionally, charge accumulation is also observed 
along the Ni-Zr/Hf bonds. The charge cloud, in this case, is more 
directional compared to the Co-containing HH and is localized closer to 
the Ni atom. Thus, in the Ni (Co) containing parent compounds, the X-Y 
bond is ionic (covalent) in nature. Interestingly, in the HEA, the 
electron rearrangement around the Ni and Co atoms remains similar to the
parent compounds with slight deviations due to local distortion in its 
structure. This shows the presence of a bonding hierarchy in the HEA. 

\subsection{Phonon dispersion and Lattice thermal conductivity}

\begin{figure}
    \centering
    \includegraphics[scale=0.28]{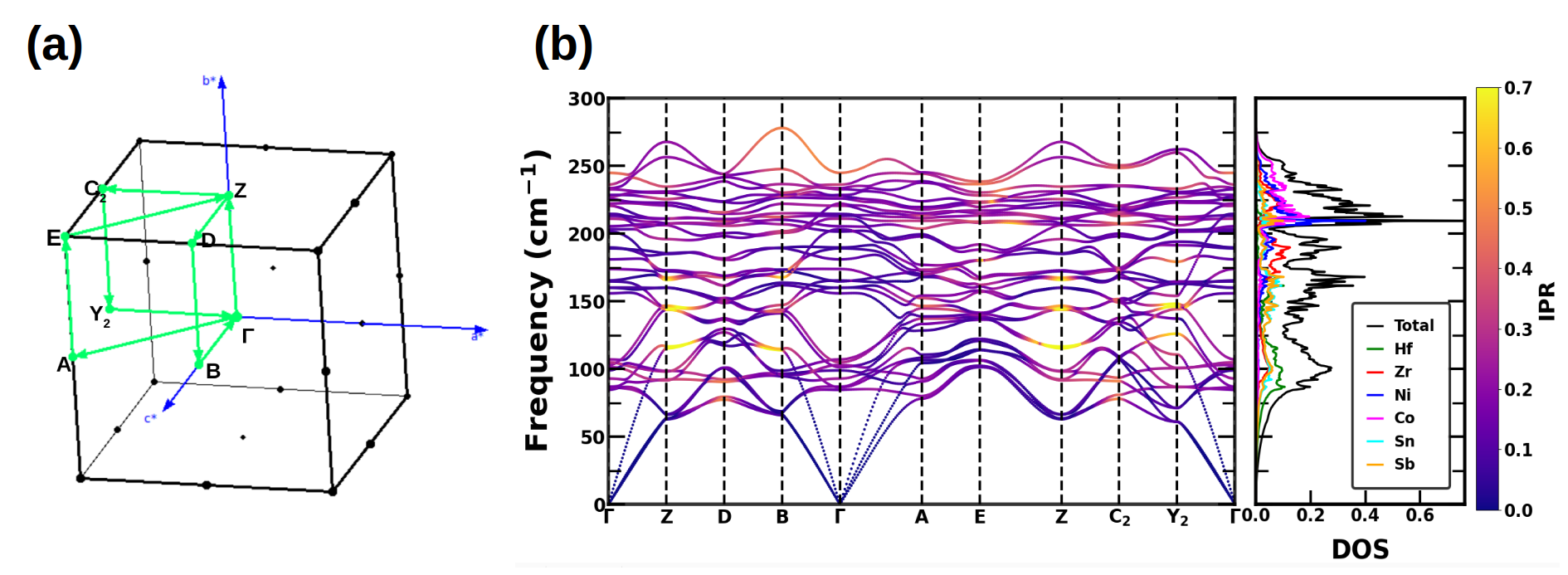}
    \caption{(a) Irreducible Brillouin zone showing the path marked in green for the HEA (b) Phonon dispersion, inverse participation ratio (IPR)
    of the phonon modes and phonon density of states of the HEA.}
    \label{fig:ph_ipr}
\end{figure}

\begin{figure}[h]
    \centering
\includegraphics[width=\columnwidth]{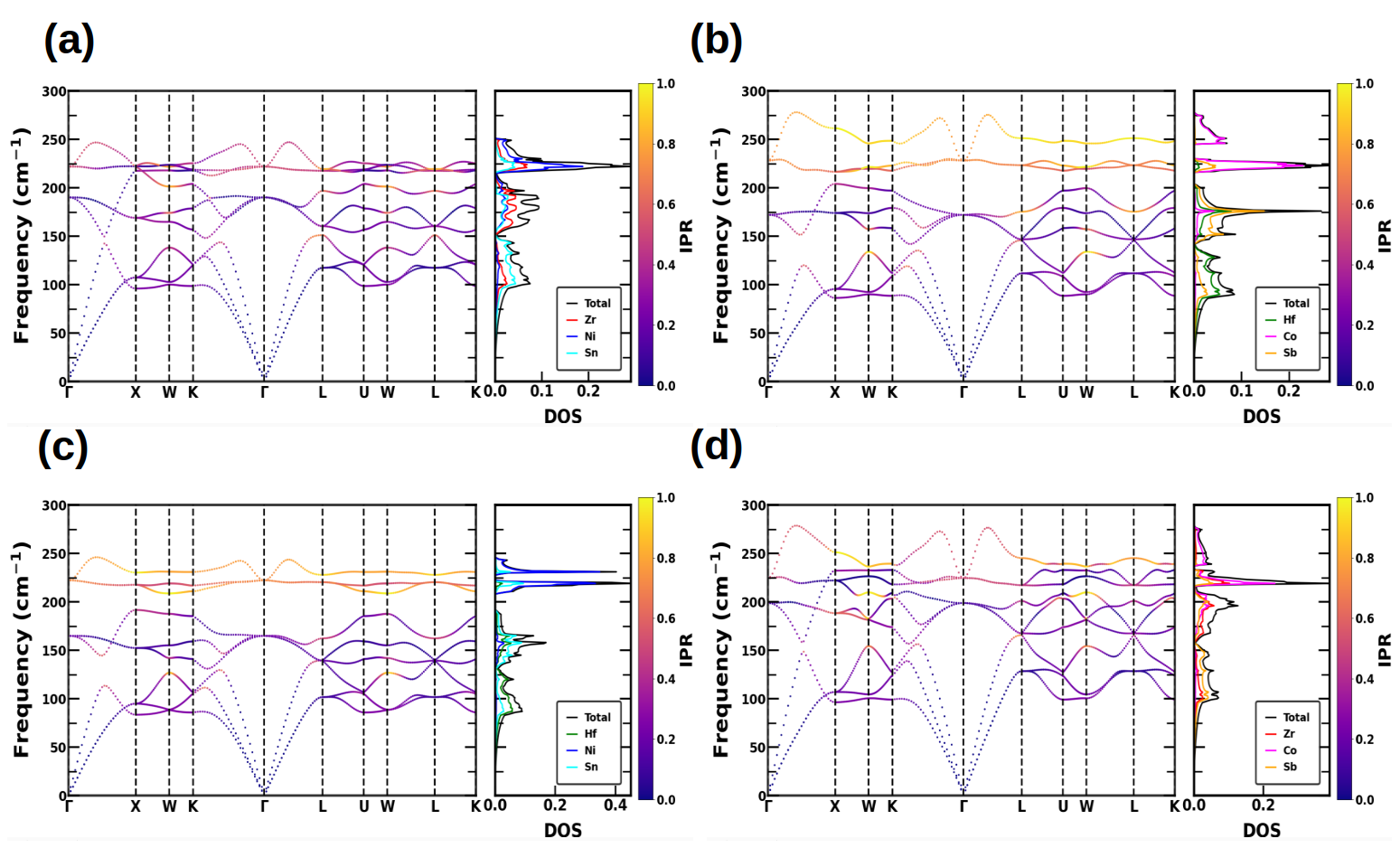} 
 	\caption{Phonon band structures, Inverse participation ratio (IPR) and phonon dos of (a) ZrNiSn, (b) HfCoSb, (c) HfNiSn, and (d) ZrCoSb }
    \label{fig:phonon_ipr_hh}
\end{figure}

In order to assess the dynamical stability of the HEA, we have computed 
its phonon spectrum, which is shown in Figure~\ref{fig:ph_ipr}. 
Further, in order to understand how the lattice vibrations are altered 
in the HEA in comparison with the parent HHs, we have also computed the 
phonon spectra of the latter (Figure~\ref{fig:phonon_ipr_hh} of the SI). The absence of 
any imaginary modes in the HEA vibrational spectrum throughout the BZ 
suggests that the HEA is dynamically stable. Compared to the parent 
HHs, where there is either nil or negligible mixing of the acoustic and 
optical phonons, the HEA spectrum shows a significant overlap of these 
phonon modes. This enhanced mixing can be attributed to the softening 
of the low frequency optical modes. Usually, this mixing between the 
heat carrying acoustic modes with the optical ones results in
scattering of the former, thereby reducing the lattice thermal 
conductivity. Moreover, in the HHs, the high frequency optical phonon 
bands are flat, giving rise to sharp peaks in the phonon density of 
states (PhDOS). In contrast, in the HEA, the optical phonon bands
become more dispersive, giving rise to the broader peaks in PhDOS.
The atom projected PhDOS show that for the modes that have frequencies 
less than 125 cm$^{-1}$ the major contribution is from the heaviest element Hf. The other heavy elements like Sb, Sn and Zr also have reasonable contributions in lower frequency modes. The Sb and Sn atoms have a dominant
contribution to the phonon modes lying between 125 cm$^{-1}$ and 175 
cm$^{-1}$. For 175 < $\omega$ < 200 cm$^{-1}$, the lattice vibrations 
are dominated by the vibrations of the Zr atoms, while those having 
frequency beyond 200 cm$^{-1}$, the major contributions are from the displacements of the lightest Ni and Co atoms.

In order to understand whether the nature of localization of the phonon 
modes in the HEA changes compared to the HH, we have computed the 
inverse participation ratio for each mode (IPR). This IPR is computed as \cite{giri2018localization} :

\begin{equation}
    IPR = \sum_{i}\left[\sum_{\alpha}\epsilon_{i\alpha ,n} \epsilon_{i\alpha , n}^{*}\right]^{2}
\end{equation}
 Here $\epsilon_{i\alpha ,n}$ represents the eigenvector component along the $\alpha$-direction for mode n. \\
\noindent $IPR=1$ ($IPR \simeq 1/N$, N being the number of atoms in the unit
cell) implies a completely localized (delocalized) phonon mode.
Figure~\ref{fig:ph_ipr}
shows the IPR for the HEA, while those of the HHs are shown in Figure~\ref{fig:phonon_ipr_hh} of SI.
While for the HHs, all the phonon modes till about 130 cm$^{-1}$
are completely delocalized, in HEA the
phonon modes with frequency greater than 50 cm$^{-1}$ tends to localize.
However, relative to the HHs where we observe that the high-frequency
phonon modes are highly localized (IPR=1 for some modes), the overall
degree of delocalization of the modes is relatively lower in the HEA
suggesting that the modes are more diffusive in nature.

\begin{figure}
    \centering
    \includegraphics[scale=0.4]{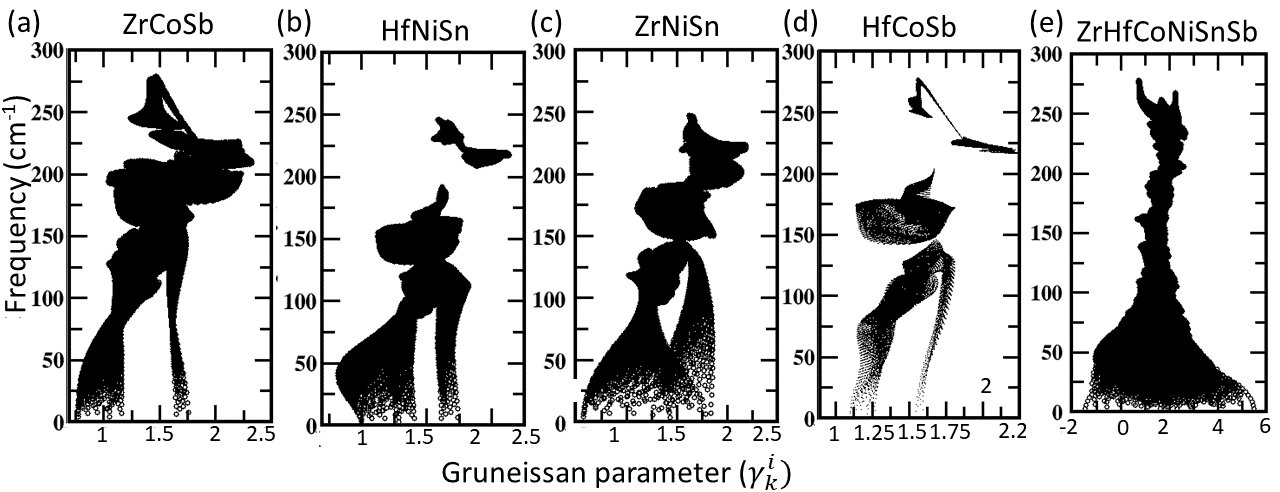}
    \caption{Mode resolved Gruneisen parameters for the parent HHs
    (a)-(d) and the HEA (e).}
    \label{fig:gp_all}
\end{figure}

The effect of hierarchical bonding, discussed in the previous section, is
reflected in the computed the mode resolved Gruneisan parameter ($\gamma_i^k$), which for the phonon of the $i^{th}$ branch with wave vector $k$ is given by:

\begin{equation}
\gamma_{i}^{k} = -\frac{V_0}{\omega_{i}^{k}}\frac{\partial \omega_{i}^{k}}{\partial V}
\label{eq:gru_mode}
\end{equation}

\noindent where $V_0$ is the equilibrium volume of the unit cell and
$\omega_{i}^{k}$ is the frequency corresponding to the phonon of the $i^{th}$ branch with wave vector $k$. The derivative in Equation~\ref{eq:gru_mode}
is evaluated numerically by using the central difference method. To achieve this, we computed the phonon spectra by applying strain, varying the lattice parameters by $\pm1\%$. The $\gamma_i^k$ of the parent HH and the HEA are 
plotted in Figure~\ref{fig:gp_all}. While for the HH $\gamma_i^k$ typically lies between 0 and 2.5, in the HEA, it varies from -2.0 to 6, 
i.e. an overall spread of 8. Moreover, this enhancement in the spread
of $\gamma_i^k$ is primarily restricted to the heat-carrying acoustic 
phonons. This suggests that the HEA lattice becomes significantly
more anharmonic compared to that of the parent HHs.

The effect of the changes in bonding of the HEA
lattice and thereby their vibrational properties also affect
thermal transport in these materials. We have computed the lattice 
thermal conductivity ($k_{L}$) for all the systems using the Debye-Callaway 
model\cite{zhang2012first, morelli2002estimation, asen1997thermal}. 
According to this model, $k_{L}$ is the sum over the contribution to
lattice thermal conductivity from one longitudinal acoustic ($k_{LA}$) and 
two transverse acoustic branches ($k_{TA}$ and $k_{TA^{\prime}}$). For the 
$i^{th}$ phonon branch (where $i = LA, TA$, or $TA^{\prime}$), the 
contribution to lattice thermal conductivity ($k_{i}$) is given by 

\begin{equation}
k_{i} = \frac{1}{3} \left(\frac{k_{B}^{4}} {{2\pi^{2}\hbar^{3}v_{i}}}\right)\left[\int_{0}^{\frac{\Theta^{i}_{D}}{T}} \frac{\tau_{c}^{i} (x) x^{4} e^{x}}{\left(e^{x}-1\right)^{2}} dx \  + \ \frac{\left(\int_{0}^{\frac{\Theta^{i}_{D}}{T}}\frac{\tau_{c}^{i}(x)x^{4}e^{4}}{\tau_{N}^{i}(x)\left(e^{x}-1\right)^2}dx\right)^{2}}{\int_{0}^{\frac{\Theta^{i}_{D}}{T}}\frac{\tau_{c}^{i}(x) x^{4}e^{x}}{\tau_{N}^{i}(x)\tau_{R}^{i}(x)\left(e^{x}-1\right)^{2}}dx}\right]
\label{eq:kappa_i}
\end{equation}

\noindent where $\hbar$ is the Plank constant, $k_{B}$ is the Boltzmann constant, $v_{i}$ is the long wavelength velocity of the $i^{th}$ mode,
and ${\tau_c^i}^{-1}$, ${\tau_N^i}^{-1}$ and ${\tau_R^i}^{-1}$ are the
scattering rates for the $i^{th}$ mode related to the total, normal
and dissipative scattering processes. $x$ in Equation~\ref{eq:kappa_i} is 
given as $x = \frac{\hbar \omega}{k_{B}T}$  where $T$ represents temperature 
and $\omega$ the phonon frequency. Further, in Equation~\ref{eq:kappa_i}, 
$\Theta^{i}_{D} $ is the Debye temperature corresponding to the $i^{th}$ 
mode. This $\Theta^{i}_{D} $ is given by \cite{zhang2016first, fan2020aicon, sahni2023double}:
		  
\begin{equation}
    \Theta^{i}_{D} = \frac{\hbar \omega_{i}^{max}}{k_{B}}
    \label{eq:thetad_i}
\end{equation}

\noindent where $\omega_{i}^{max}$ is the maximum phonon frequency for the 
$i^{th}$ mode.

\noindent The $\tau_c$, $\tau_N$ and $\tau_R$ in Equation~\ref{eq:kappa_i} are related 
as:
\begin{equation}
    \tau_{c}^{-1} = \tau_{N}^{-1} + \tau_{R}^{-1}
    \label{eq:tau_t}
\end{equation}
\noindent For normal phonon scattering, the corresponding relaxation time for
the longitudinal ($\tau^{LA}_{N}$) and transverse acoustic 
($\tau_{N}^{TA}/\tau_{N}^{TA^{^{\prime}}}$) modes: 
		  
\begin{equation}
    \frac{1}{\tau^{LA}_{N}(x)} \  = \ \frac{k_{B}^{3}\gamma_{LA}^{2}V_{a}}{M_{a}\hbar^{2}v_{LA}^{5}}\left(\frac{k_{B}}{\hbar}\right)^{2}x^{2}T^{5}
    \label{eq:tau_la}
\end{equation}

\begin{equation}
   \frac{1}{\tau^{TA/TA^{'}}_{N}} =\frac{k_{B}^{4}\gamma_{TA/TA^{'}}^{2}V_{a}}{M_{a}\hbar^{3}v_{TA/TA^{'}}^{5}}\left(\frac{k_{B}}{\hbar}\right)xT^{5}
   \label{eq:tau_ta}
\end{equation}

\noindent where $\gamma_i = \sqrt{<\left(\gamma_{i}^{k}\right)^{2}>} $ is
the mode averaged Gruneissan parameter, $M_a$ is the  average atomic mass per unit cell
and $V_a$ is the volume per atom.

In most crystalline solids, the dissipative scattering is primarily due to 
Umklapp processes and the corresponding relaxation time ($\tau_{U}^{i}$) for 
the $i^{th}$ mode is given by :
		  
\begin{equation}
   \frac{1}{\tau_{U}^{i}(x)} =  \frac{\hbar\gamma^{2}}{M_{a}v_{i}^{2}\Theta_{D}^{i}} \left(\frac{k_{B}}{\hbar}\right)^{2}x^{2} T^{3} e^{-\frac{\Theta_{D}^{i}}{3T}}
   \label{eq:tau_u}
\end{equation}

\noindent Hence, for the parent HHs, for $i^{th}$  mode, the total phonon 
scattering rate (${\tau_c^{-1}}^{HH,i}$) depends on the scattering rates 
associated with the normal and the Umklapp processes and is given by:

\begin{equation}
     \frac{1}{\tau_{c}^{HH,i}} = \frac{1}{\tau_{N}^{i}} \ + \frac{1}{\tau_{U}^{i}}
     \label{n_um}
\end{equation}

\noindent However, in the case of ZrHfCoNiSnSb  HEA, significant mass 
fluctuation occurs at the Wyckoff position 4b (0.5, 0.5, 0.5) which
can now be occupied either by Hf or Zr, the latter having a mass half of
that of Hf. Further, the 4a (0, 0, 0) position which is now occupied by
either Sb or Sn will also exhibit mass fluctuations due to their different
atomic masses. Hence, we expect that the propagating phonons will be
scattered also by these mass defects in HEA caused by these mass fluctuations 
plays a crucial role in phonon scattering. Hence, to compute the
scattering rates for dissipative processes in the HEA, we have also 
incorporated the effect of mass fluctuation scattering. Using Klemens\cite{klemens1960thermal} formalism, the relaxation time for mass fluctuation scattering ($\tau_{M}^{i}$) is given by:

\begin{equation}
\frac{1}{\tau_{M}^{i}} = \left(\frac{V_{a}k_{B}^{4}}{4\pi\hbar^{4}v_{i}^{3}}\right)x^{4}T^{4}\Gamma_{M}
\label{eq:tau_m}
\end{equation}

\noindent where the disordered scattering parameter $\Gamma_{M}$ is given
by :
		  
\begin{equation}
 \Gamma_{M} \ = \ \frac{\sum_{j=1}^{n}c_{j}\left(\frac{\overline{M_{j}}}{\overline{\overline{M}}}\right)^{2}f_{j}^{1}f_{j}^{2}\left(\frac{M_{j}^{1}-M_{j}^{2}}{\overline{M_{j}}}\right)^{2}} {\sum_{j=1}^{n}c_{j}}
 \label{eq: g_m}
\end{equation}
		  
\noindent where $c_{j}$ represents the relative site degeneracy, $f_{j}$ 
denotes the fractional occupation, $\overline{M_{j}}$ is the average mass at 
the site $j$, and $\overline{\overline{M}}$ is the average atomic mass of the 
compound. We note that these corrections have been successfully applied
to double HHs previously\cite{petersen2015critical, fan2020aicon}. 
Consequently, the total relaxation time for the $i^{th}$ mode of ZrHfCoNiSnSb is given by : \\

\begin{equation}
\frac{1}{\tau_{c}^{HEA,i}} = \frac{1}{\tau_{N}^{i}} \ + \frac{1}{\tau_{U}^{i}} + \frac{1}{\tau_{M}^{i}}
\label{eq:tau_full_hea}
\end{equation}

The values of the different parameters used to compute $k_L$ in
Equation~\ref{eq:kappa_i} are given in Table~\ref{table:vib_prop}.

\begin{table} [ht]
\centering 
\caption{Mode resolved Debye temperature ($\Theta_D^i$), long wavelength
phonon velocity ($v_i$) and Gruneisen parameter ($\gamma_i$) for the
parent systems and the HEA. The values given in square brackets are
from Ref.~\cite{sahni2023double})}
\begin{tabular}{|c|c|c|c|c|c|}
\hline
Property & ZrNiSn & HfCoSb &  HfNiSn & ZrCoSb & ZrHfCoNiSnSb    \\
\hline 
$\Theta_D^{LA}$ (K) & 217 [214] & 210 & 201 [196] & 238 & 153 \\
\hline
$\Theta_D^{TA}$ (K) & 170 [166] & 161 & 147 [145] & 185 & 146 \\
\hline
$\Theta_D^{TA^{'}}$ (K) & 186 [184] & 171 & 160 [160] & 200 & 147 \\
\hline
$v_{LA}$ (m/s) & 5338 [5323] & 5115 & 4830 [4746] & 5746 & 5224 \\
\hline
$v_{TA}$ (m/s) & 3080 [2852] & 2843 & 2540 [2508] & 3115 & 2499 \\
\hline
$v_{TA^{'}}$ (m/s) & 3080 [3600] & 2843 & 2540 [3120] & 3115 & 2683 \\
\hline	
$\gamma_{LA}$ & 1.47 [1.69] & 1.67 & 1.54 [1.55] & 1.49 & 1.42 \\
\hline
$\gamma_{TA}$ & 1.22 [1.30]  & 1.36 & 1.28 [1.35] & 1.23 & 1.41 \\
\hline
$\gamma_{TA^{'}}$ & 1.28 [1.36] & 1.43 & 1.41 [1.38] & 1.24 & 1.60 \\
\hline
$\gamma$ & 1.61 [1.71] & 1.64 & 1.59 [1.59] & 1.59 & 1.70 \\
\hline
\end{tabular}
\label{table:vib_prop}
\end{table}

\begin{figure}
    \centering
    \includegraphics[scale=0.42]{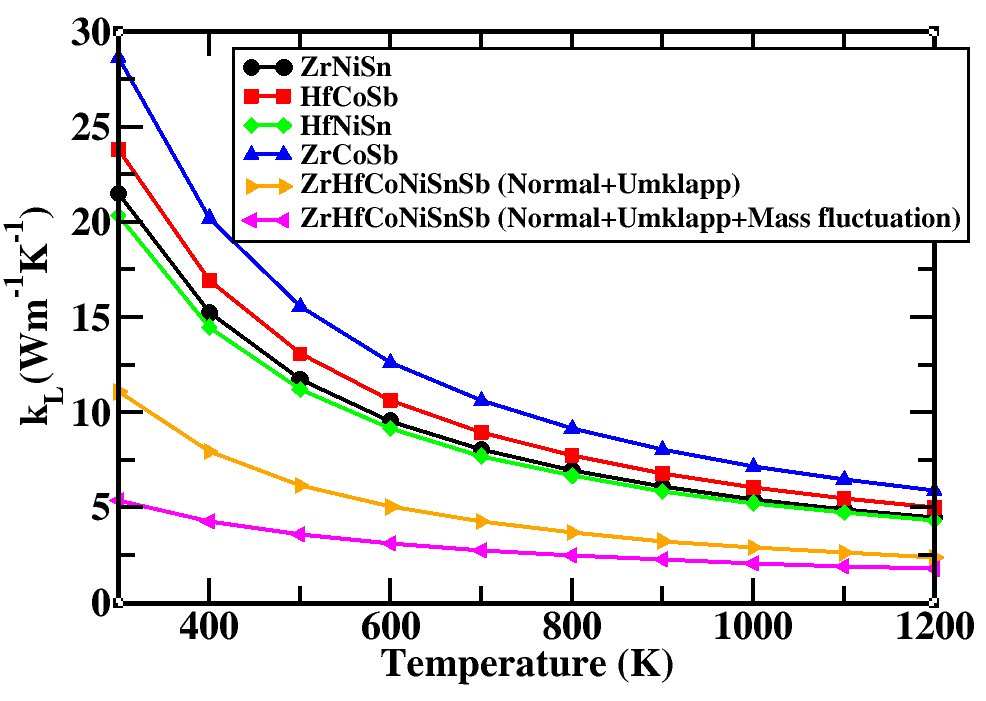}
    \caption{Lattice thermal conductivity as a function of temperature.}
    \label{fig:kappa_l}
\end{figure}

Figure~\ref{fig:kappa_l} shows the lattice thermal conductivity of the HHs and the HEA.
We observe that at 300 K the $k_L$ of the HHs lie between 20.36 and 28.64 Wm$^{-1}$K$^{-1}$
with HfNiSn (ZrCoSb) having the lowest (highest) value. We note
that our computed values of $k_L$ is in reasonably good agreement
with those reported in the literature using the solutions of the semiclassical
Boltzmann transport equations for phonons that are more
computationally demanding but accurate \cite{carrete2014finding, gandi2017thermoelectric, gandi2016electron}.
For all the systems, the lattice thermal conductivity is reduced
with an increase in temperature.
For the HEA, when we compute the lattice thermal conductivity by including
only the Umklapp scattering, we obtain a value of 11.09 Wm$^{-1}$K$^{-1}$ at 300 K. We note
that this is less than half of the values observed in the HHs. This drastic
reduction in lattice thermal conductivity can be attributed to the 
different types of chemical bonding observed in the HEA lattice that resulted in enhanced anharmonicity. On incorporation
of the scattering effects due to mass fluctuations, the lattice thermal
conductivity is further reduced to 5.39 Wm$^{-1}$K$^{-1}$ at 300 K. Thus, our results
suggest that the synergistic effect of the changes in bonding in the HEA
lattice and the mass fluctuations can drastically reduce $k_L$. At room temperature, the $k_{L}$ of the HEA is reduced by a factor of three compared to the parent HHs Hf/ZrNiSn, and by a factor of five compared to Zr/HfCoSb. 

\subsection{Electronic properties}

\begin{figure}
    \centering
\includegraphics[width=\columnwidth]{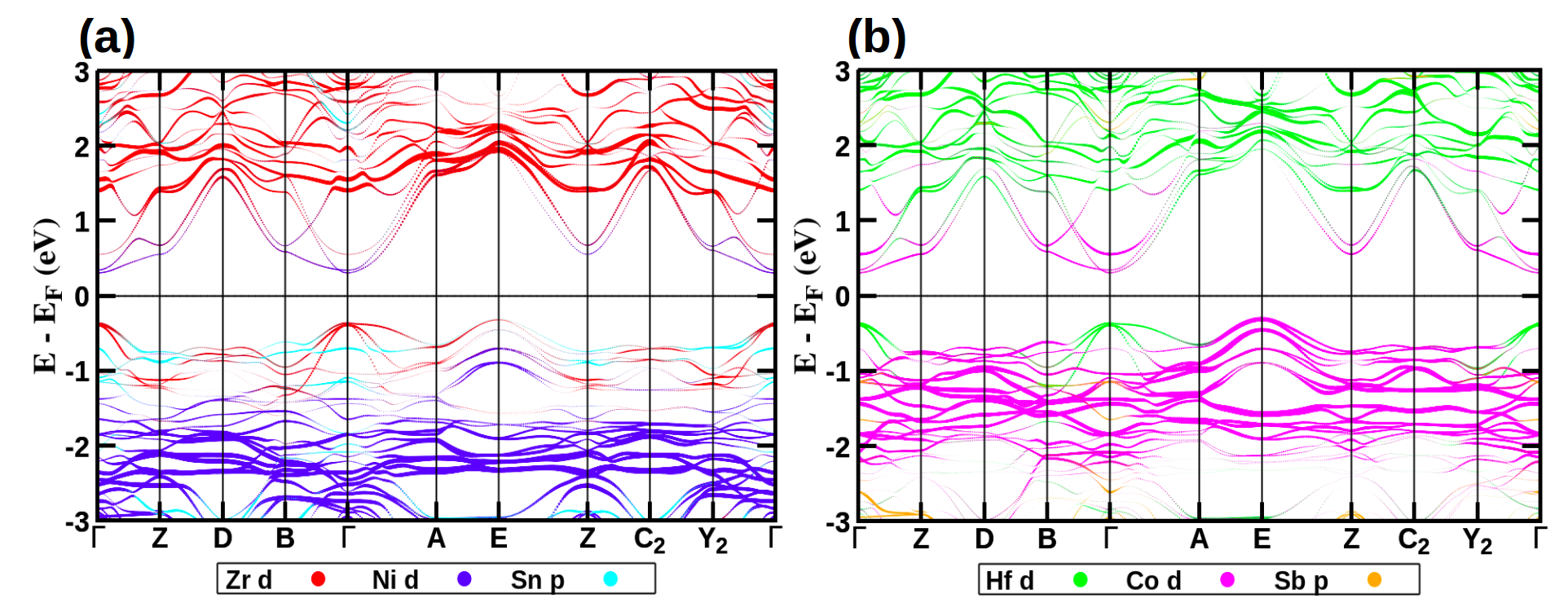} 
    \caption{ Band structure of HEA showing contributions from (a) Zr-d (red), Ni-d (blue), and Sn-p (turquoise) orbitals (b) Hf-d (green), Co-d (pink), and Sb-p (orange) orbitals.}
    \label{fig:hea_bs}
\end{figure}

\begin{figure}[h]
    \centering
\includegraphics[width=\columnwidth]{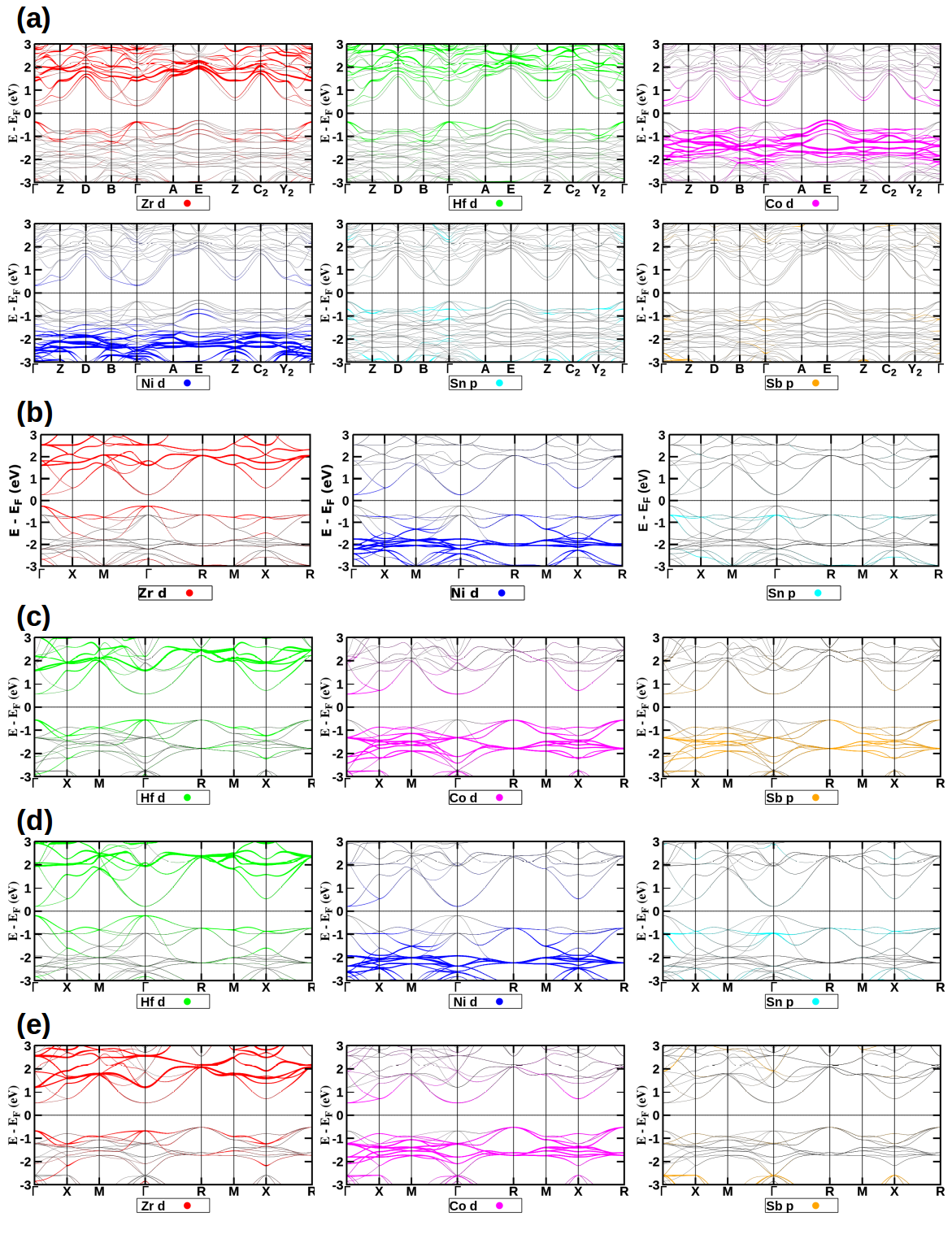} 
    \caption{Atomic orbital contributions to the band structure for (a) ZrHfCoNiSnSb  (b) ZrNiSn (c) HfCoSb (d) HfNiSn (e) ZrCoSb}
    \label{fig:orbital_contri}
\end{figure}

Figure~\ref{fig:hea_bs} shows
the band structure of ZrHfCoNiSnSb, along with the contributions from the
d-states of the transition metals and the p-states of the p-block
elements. Those for the parent HHs are shown in Figure~\ref{fig:orbital_contri} of the SI.
In accordance with the literature report, we observe that all the HHs are
semiconducting in nature, with HfCoSb having the largest band gap of
1.12 eV \cite{gandi2017thermoelectric}, followed by 1.05 eV for ZrCoSb \cite{gandi2017thermoelectric}, 0.51 eV for 
ZrNiSn \cite{gandi2016electron}  and 0.40 eV for HfNiSn \cite{gandi2016electron} . All these HHs have the 
conduction band minima (CBM) at the
$\Gamma$ point of the BZ. However, their valence band maxima (VBM) occurs
at different points of the BZ. Similar to the CBM, the VBM of ZrNiSn and 
HfNiSn is at the $\Gamma$ point of the BZ, making these two HHs as a
direct band gap semiconductor. In contrast, ZrCoSb has VBM at the R point of
the BZ making it an indirect band gap semiconductor. In the case of HfCoSb,
there are two degenerate VBMs, namely at $\Gamma$ and the R point of the BZ.
Additionally, the VB and CB edges of Zr/HfNiSn are closer to the Fermi
energy than those of Zr/HfCoSb.

Similar to the HHs, the CBM of the HEA is at the $\Gamma$ point of the
BZ while the VBM is at E-point (Figure~\ref{fig:hea_bs}). Since, the 
lattice parameters of this monoclinic lattice are very close and the deviation 
of the $\gamma$ from 90$^{\circ}$ is negligibly small, the E point of the BZ 
of this monoclinic lattice is the same as that of the R-point of the BZ of the 
cubic lattice. The HEA is an indirect semiconductor with a band gap of
0.61 eV, which is less than that of the Zr/HfCoSb and more than that of
Zr/HfNiSn. Additionally, the maxima of the valence band at the $\Gamma-$point 
is only 60 meV below the VBM, suggesting that at high temperatures,
p-type carriers belonging to this hole pocket will also contribute to the
transport properties. We note that these features of the valence band are
very similar to those observed in ZrCoSb (Figure~\ref{fig:orbital_contri} (e) of SI) and HfCoSb
(Figure~\ref{fig:orbital_contri}(c) of SI). Furthermore, the conduction band of the HEA have additional electron pockets at the B and Z points of the BZ. These are
about 280 and 250 meV higher in energy compared to the CBM. These
high symmetry points of the BZ of the HEA are analogous to the X-point of the
cubic BZ of the HHs. While the VBM at E-point have contributions from
Co-d states, the hole pocket at the $\Gamma$ point has a contribution
from Zr and Hf-d states (Figure~\ref{fig:hea_bs} and Figure ~\ref{fig:orbital_contri} of SI).
The Ni-d states lie deep inside the valence band. This is in contrast
to that observed in Hf/ZrNiSn, where Ni-d orbitals contributed to hole pockets
at the R-point of the cubic BZ. The CBM at $\Gamma$ and the electron
pockets at B and Z-points in the BZ of the HEA have contributions from
the d-orbitals of all the transition metal elements. 

\subsection{Conductivity and density of states effective masses}
\label{sec:eff_mass}

Effective masses ($m^{*}$) of charge carriers, which is a key component of 
transport properties like conductivity, mobility, Seebeck coefficient, etc., 
had been computed at the different valence and conduction band extrema within
the parabolic band approximation. $m^{*}$ is given by:
       	 
\begin{equation}
    \frac{1}{m^{*}} = \frac{1}{\hbar ^{2}} \frac{\partial^{2}\epsilon}{\partial k ^{2}}
    \label{eq:eff_mass}
\end{equation}
       	 
\noindent Near the extrema, the isosurfaces of energy are ellipsoids, and hence 
the effective mass is different along the longitudinal or transverse 
directions. The conductivity effective mass, which affects relaxation
time and electrical conductivity, $m_{\sigma}^{*}$, is given by the harmonic
mean of the longitudinal effective mass ($m^*_{l}$) and the two transverse
effective masses (${m^*_{t1}}$ and ${m^*_{t2}}$) as:

\begin{equation}
    \frac{1}{m_{\sigma}^{*}} = \frac{1}{3}\left(\frac{1}{m^*_{l}}  + \frac{1}{m^*_{t1}} + \frac{1}{m^*_{t2}} \right).
    \label{eq:m_sigma}
\end{equation}

\noindent Similarly, the density of states (DOS) effective mass $m_{D}^{*}$ is 
given by the geometric mean of the three masses weighted by the 
       	 
\begin{equation}
    m_{D}^{*} = N_{v}^{\frac{2}{3}} \left(m_{l}.m_{t_{1}}.m_{t_{2}}\right)^{\frac{1}{3}}
    \label{eq:m_dos}
\end{equation}

The conductivity and DOS effective masses at the different extrema of the
BZ are listed in Tables~\ref{table:eff_mass_cbm_hh} to \ref{table:eff_mass_vbm_hea} of SI for the normal HHs and the HEA.

\subsection{Carrier relaxation times}
\label{sec:tau}

The relaxation times ($\tau$) of the charge carriers in these materials
were calculated using the deformation potential 
theory\cite{bardeen1950deformation}, which takes into account of their scattering by the acoustic phonons only. The relaxation time of carriers in a 
band $b$ having dos effective mass $m_{D}^{*}$ is given by: 
       	 
\begin{equation}
    \tau_{b} = \frac{2\left(2\pi\right)^{\frac{1}{2}} \hbar^{4}C }{3 \Xi^{2} \left(k_{b}T \right)^{\frac{3}{2}} \left(m_{D}^{*}\right)^{\frac{3}{2}}  }
    \label{eq:tau_b}
\end{equation}

\noindent In the above equation, elastic constant $C$ and deformation 
potential $\Xi$ are computed as: 
       	 
\begin{equation}
    C = \left(\frac{1}{V_{0}} \frac{\partial ^{2}E}{\partial \left(\frac{\Delta a }{a_{0}}\right)^{2}}\right)_{a=a_{0}} \\\ \ \ \ \ , \ \ \ \ \ \ \   \Xi = \left(\frac{\partial E_{edge}}{\partial \left(\frac{\Delta a}{a_{0}}\right)}\right)_{a=a_{0}}
    \label{eq:c_xi}
\end{equation}

\noindent Here $E$ is the total energy of the system obtained from DFT
calculation, $a_{0}$ is the optimized lattice constant, $\Delta a = a-a_{0}$ 
is the lattice distortion from its equilibrium value, $V_{0}$ is the 
equilibrium volume of the unit cell, and $E_{edge}$ is the energy of VB or 
CB extrema. The carrier mobility corresponding to this band $\mu_{b}$ is 
expressed as: 
       	 
\begin{equation}
    \mu _{b} = \frac{e\tau _{b}}{m_{\sigma}^{*}}
    \label{eq:avg_mob}
\end{equation}
       	 
\noindent When there are many bands that are either degenerate or near degenerate to valence band (VB) or conduction band (CB), average carrier mobility $\mu_{av}$ and average conductivity effective mass $m_{\sigma,av}^{*}$ are given by\cite{liu2019carrier}:
       	 
\begin{equation}
    \mu_{av} = \sum_{b} \frac{n_{b}}{n} \mu_{b} \ \ ,  \ \ \ \frac{1}{m_{\sigma, av}^{*}} = \sum_{b} \frac{n_{b}}{n} \frac{1}{m_{\sigma,b}^{*}}
    \label{eq:msigma_avg}
\end{equation}
       	 
\noindent In the above equations, $n_{b}$ is the number of charge carriers in 
the valley of the band $b$ and $n = \sum_{b} n_{b}$ is the total number 
of the charge carriers. The fraction of the charge carriers carried by the 
valley of the band $b$ and VB/CB is given by 

\begin{equation}
    \frac{n_{b}}{n_{VB/CB}} = \left(\frac{m_{D, b}^{*}}{m_{D,VB/CB}^{*}}\right)^{\frac{3}{2}} \exp \left(-\frac{\Delta E}{k_{b}T}\right)
    \label{eq:relative_carr_conc}
\end{equation}  
       	 
\noindent where $\Delta E$ is the difference in energy between the valley 
extrema and VBM/CBM. Finally, the average relaxation time, which considers the 
contributions from all the valley extrema are given by\cite{liu2019carrier}:
       	 
\begin{equation}
    \tau_{av} = \frac{m_{\sigma, av}^{*}\mu_{av}}{e}
    \label{eq:tau_avg}
\end{equation}  

\noindent The quantities required to compute $\tau_{av}$ using the above
equations and the values of $\tau_{av}$ and $\mu_{av}$ for electrons and holes at 300 K for
all the compounds are given in Tables~\ref{table:tau_e} and \ref{table:tau_h}, respectively. It is observed that all the parent
HHs, except HfCoSb, the average conductivity effective mass of
electrons is greater than that of holes. In contrast, for the HEA HHs
$m_{\sigma, av}^{*}$ of electrons are smaller than that of holes.
Further, we observe that the magnitude of the deformation potential
of electrons and holes, 
which is a measure of the coupling between the charge carriers and acoustic phonons are similar. However, we find that $|\Xi|$ of electrons
for all the compounds are larger than that observed in holes. This
suggests that in these materials, the electron-acoustic phonon
coupling is larger than that between holes and acoustic phonons.

\begin{table} [ht]
\centering 
\caption{Average conductivity effective mass ($m_{\sigma, av}^{*}$), deformation potentials ($|\Xi|$),
elastic constants ($C$), average mobility ($\mu_{av}$)and average relaxation time ($\tau_{av}$) for
electrons in the different materials. The values of $\mu_{av}$ and
$\tau_{av}$ reported in the table had been computed at 300 K.}
\begin{tabular}{|c|c|c|c|c|c|}
\hline
Properties  & ZrNiSn & HfCoSb &  HfNiSn & ZrCoSb & ZrHfCoNiSnSb   \\
\hline 
$m_{\sigma, av}^{*}$& 2.13 & 4.37 & 2.10 & 3.46 & 0.92 \\
\hline
$|\Xi|$ (eV) & 16.12 & 15.51 & 16.20 & 15.42 & 15.75 \\
\hline
$C$ (GPa) & 233 & 275 & 242 & 264 & 247 \\
\hline
$\mu_{av}$ (cm$^{2}$/Vs) & 42.8 & 6.65 & 48.17 & 7.47 & 60.02 \\
\hline
$\tau_{av}$ (fs) & 51.73 & 16.52  & 57.61& 14.72 &  31.45 \\	
\hline
\end{tabular}
\label{table:tau_e}
\end{table}
   
\begin{table} [ht]
\centering 
\caption{Average conductivity effective mass ($m_{\sigma, av}^{*}$), 
deformation potentials ($|\Xi|$), elastic constants ($C$), average mobility 
($\mu_{av}$)and average relaxation time ($\tau_{av}$) for holes in the 
different materials. The values of $\mu_{av}$ and $\tau_{av}$ reported in 
the table had been computed at 300 K.}
\begin{tabular}{|c|c|c|c|c|c|}
\hline
Properties  & ZrNiSn & HfCoSb &  HfNiSn & ZrCoSb & ZrHfCoNiSnSb    \\
\hline 
$m_{\sigma, av}^{*}$ & 0.75 & 5.74 & 0.61 & 1.54 & 1.27 \\
\hline
$|\Xi|$ (eV) & 15.51 & 14.82 & 15.70 & 14.84 & 15.17 \\
\hline
$C$ (GPa) & 233 & 275 & 242 & 264 & 247  \\
\hline
$\mu_{av}$ (cm$^{2}$/Vs) & 153.3 & 8.33 & 241.68 & 3.91 & 5.30 \\
\hline
$\tau_{av}$ (fs) & 65.23 & 27.20 &83.61 & 3.41 & 3.83 \\    
\hline
\end{tabular}
\label{table:tau_h}
\end{table}

\begin{figure}[h!]
\centering
\includegraphics[width=0.90\textwidth]{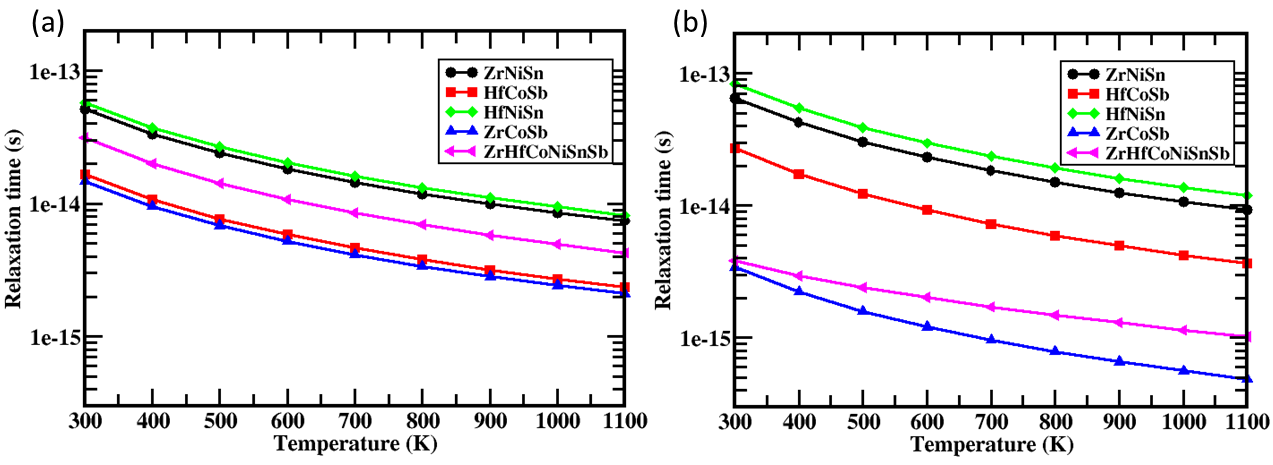}
\caption{ Average relaxation time ($\tau_{av}$) of (a) electrons
and (b) holes in the parent HHs and the HEA.}
\label{fig:tau_avg_eh}
\end{figure}
		
Figure~\ref{fig:tau_avg_eh} shows the variation of $\tau _{av}$ as a function of T. 
We find that $\tau_{avg}$ of electrons in the HEA is smaller (larger)
than that in ZrNiSn and HfNiSn (ZrCoSb and HfCoSb). For the holes,
we find that while $\tau_{avg}$ of the HEA is larger than that
observed in ZrCoSb, it is smaller than that observed in the other 
parent HHs, namely ZrNiSn, HfCoSb and HfNiSn.

	   \subsection{Electronic transport calculation}

\begin{figure}[h!]
    \centering
\includegraphics[width=\columnwidth]{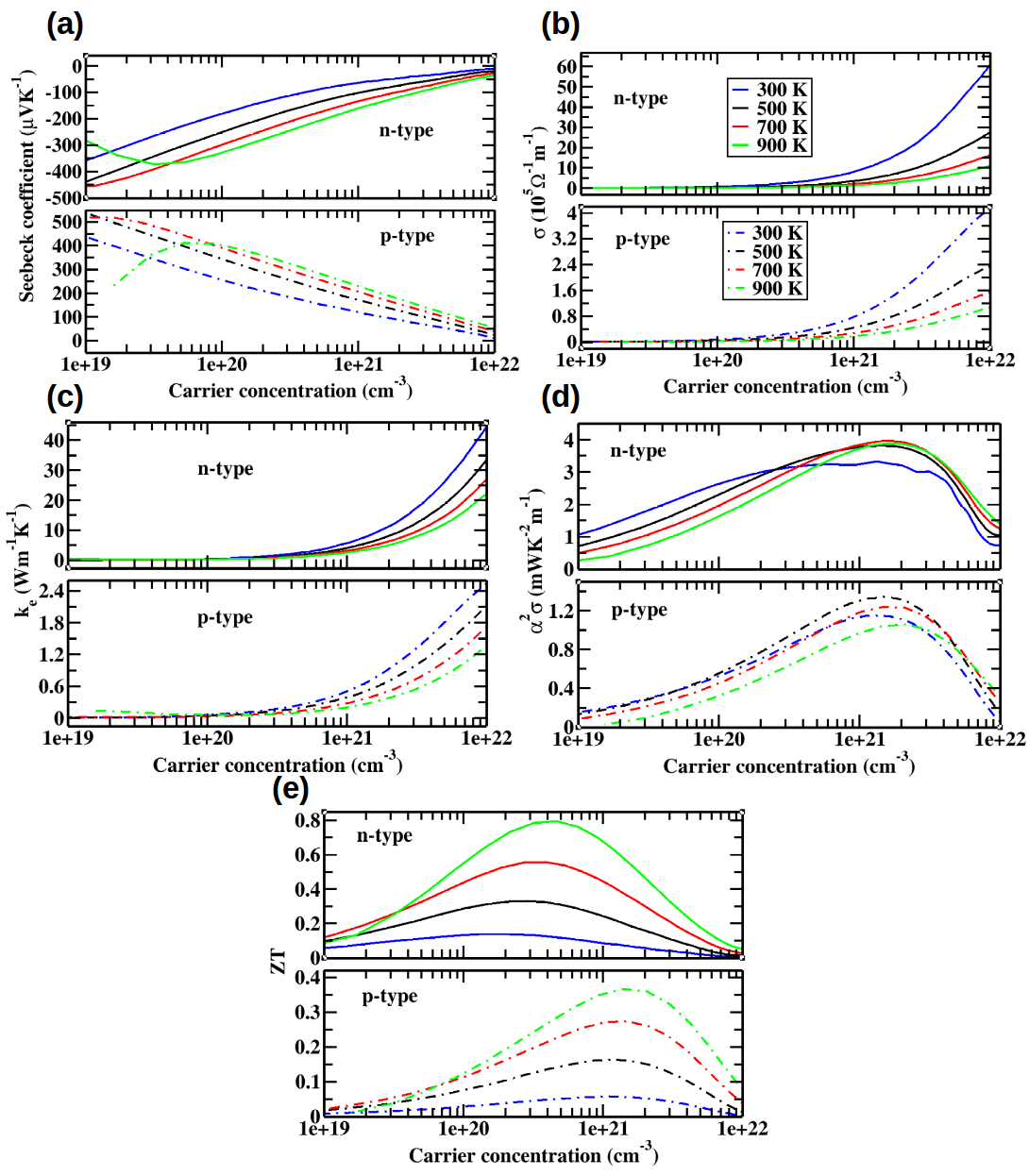} 
		\caption{(a) Seebeck coefficient, (b) Electrical conductivity, (c) Electronic thermal conductivity, (d) Power factor and (e) Figure of merit as a function of carrier concentration at different temperatures for ZrHfCoNiSnSb}
 \label{fig:electronic_properties}
\end{figure}

     Full electronic transport calculations were performed by combining the Boltzmann transport equation with a constant relaxation time approximation and a relaxation time derived from the deformation potential theory. Figure~\ref{fig:electronic_properties} (a)-(d) illustrates the variation in the electronic transport properties of ZrHfCoNiSnSb with carrier concentration, ranging from $10^{19}$ to $10^{22}$ cm$^{-3}$, at different temperatures. The Seebeck coefficient exhibits an initial increase followed by a decrease at elevated temperatures as the carrier concentration increases. With increasing temperature, the peak position shifts towards higher carrier concentrations while the peak magnitude decreases. At 900 K, the maximum Seebeck coefficient reaches 372 $\mu$V/K for n-type carriers and 413 $\mu$V/K for p-type carriers, occurring at carrier concentrations of $3.19 \times 10^{19}$ cm$^{-3}$ and $5.29 \times 10^{19}$ cm$^{-3}$, respectively. These concentrations correspond to chemical potentials of 0.16 eV below the conduction band minimum (CBM) for n-type carriers and 0.22 eV above the valence band maximum (VBM) for p-type carriers. \\
 Figure~\ref{fig:electronic_properties}(b) presents the electrical conductivity ($\sigma$) as a function of carrier concentration at different temperatures for both n-type and p-type carriers. The plot reveals that $\sigma$ for n-type carriers is higher than that for p-type carriers. This is evident from the fact that the average effective conductive mass of electrons is smaller than that of holes, as indicated in Tables~\ref{table:tau_e} and \ref{table:tau_h}. The electrical conductivity for both types of carriers remains low up to a carrier concentration of approximately $10^{20}$ cm$^{-3}$. This behavior arises because the chemical potential remains within the bandgap in this carrier concentration range, leading to a negligible value of the projected conductivity tensor $\sigma_{ij}(\epsilon)$, as defined by Equation~\ref{eq:proj_s}, at the peak of the selection function $\phi = -\partial f_{0}(T;\mu)/ \partial \epsilon$. Consequently, only the tail of the selection function contributes to the electrical conductivity, resulting in extremely low values of $\sigma$. Beyond a carrier concentration of approximately $10^{20}$ cm$^{-3}$, the electrical conductivity increases with increasing carrier concentration. This suggests that the chemical potential has shifted into the conduction or valence bands, where $\sigma_{ij}(\epsilon)$ makes a significant contribution to the overall electrical conductivity $\sigma_{ij}(T;\mu)$ at the peak of the selection function. According to the Wiedemann-Franz law, the electronic contribution to the thermal conductivity ($k_{e}$) follows a similar trend as $\sigma$, as depicted in Figure~\ref{fig:electronic_properties} (c). \\
The performance of a thermoelectric device is characterized by the power factor ($S^{2}\sigma$), which directly influences its efficiency. As shown in Figure~\ref{fig:electronic_properties} (a), the Seebeck coefficient generally decreases with increasing carrier concentration across most of the concentration range. Meanwhile, Figure~\ref{fig:electronic_properties}(b) indicates that electrical conductivity increases consistently throughout the entire carrier concentration range. Consequently, the power factor exhibits an optimal value, as observed in Figure~\ref{fig:electronic_properties}(d). With increasing temperature, the peak position of the power factor shifts toward higher carrier concentrations. 
From Figure~\ref{fig:electronic_properties}(d), it is evident that for both n-type and p-type carriers, the peak power factor initially increases and then decreases with rising temperature. For the n-type case, ZrHfCoNiSnSb achieves a maximum power factor of 3.32 (3.88) mW K$^{-2}$ m$^{-1}$ at a carrier concentration of $1.26 \times 10^{21}$ ($1.54 \times 10^{21}$) cm$^{-3}$ at 300 (900) K, for the chemical potential is located 0.24 (0.23) eV above the conduction band minimum (CBM). In contrast, for the p-type case, the maximum power factor reaches 1.16 (1.06) mW K$^{-2}$ m$^{-1}$ at a carrier concentration of $1.37 \times 10^{21}$ ($2.08 \times 10^{21}$) cm$^{-3}$ at 300 (900) K, corresponding to a chemical potential of 0.13 (0.09) eV below the valence band maximum (VBM).
For comparison, the transport properties of different systems were plotted against the carrier concentration at 900 K in the Supplementary Information (SI). As shown in Figure~\ref{fig:ep_comp_ec}(a), which depicts electrical conductivity, HfNiSn exhibits the highest electrical conductivity across all carrier concentrations for the n-type case, followed by ZrNiSn, ZrHfCoNiSnSb, HfCoSb, and ZrCoSb. Similarly, for the p-type case, HfNiSn again shows the highest electrical conductivity, followed by ZrNiSn, HfCoSb, ZrHfCoNiSnSb, and ZrCoSb. Figure~\ref{fig:ep_comp_seebeck}(a) illustrates the variation of the Seebeck coefficient as a function of carrier concentration at 900 K for different structures. For the n-type case, ZrCoSb exhibits the highest Seebeck coefficient, followed by HfCoSb, ZrHfCoNiSnSb, ZrNiSn, and HfNiSn. In the p-type case, HfCoSb has the highest Seebeck coefficient, followed by ZrCoSb, ZrHfCoNiSnSb, ZrNiSn, and HfNiSn. These trends in electrical conductivity and the Seebeck coefficient are explained using the projected conductivity tensor $\sigma_{ii}(\epsilon)$, the selection function $\phi = - \partial f_{0}(T;\mu)/\partial \epsilon$, and the projected Seebeck tensor $\alpha_{ii}(\epsilon)$ (as defined in Equation~\ref{eq:projected_seebeck_tensor}) in the Supplementary Information. Figure~\ref{fig:other_trans_prop}(a) in SI presents the variation of the electronic component of thermal conductivity as a function of carrier concentration for different structures at 900 K. In accordance with the Wiedemann-Franz law, this electronic contribution to thermal conductivity follows a similar trend to that of $\sigma$, with the structural hierarchy remaining consistent. Figure~\ref{fig:other_trans_prop}(b) illustrates the variation of the power factor as a function of carrier concentration for different structures. As observed in the plot, HfNiSn and ZrNiSn exhibit the highest power factor for the n-type case, primarily due to their high electrical conductivity, followed by ZrHfCoNiSnSb, HfCoSb, and ZrCoSb. Similarly, for the p-type case, HfNiSn and ZrNiSn again demonstrate the highest power factor, followed by HfCoSb. However, for the p-type case, ZrHfCoNiSnSb and ZrCoSb show a power factor approximately an order of magnitude lower than the other systems, predominantly due to their lower electrical conductivity.

	 \bigskip

\subsection{Figure of merit}

\begin{figure}[h!]
    \centering
\includegraphics[width=\columnwidth]{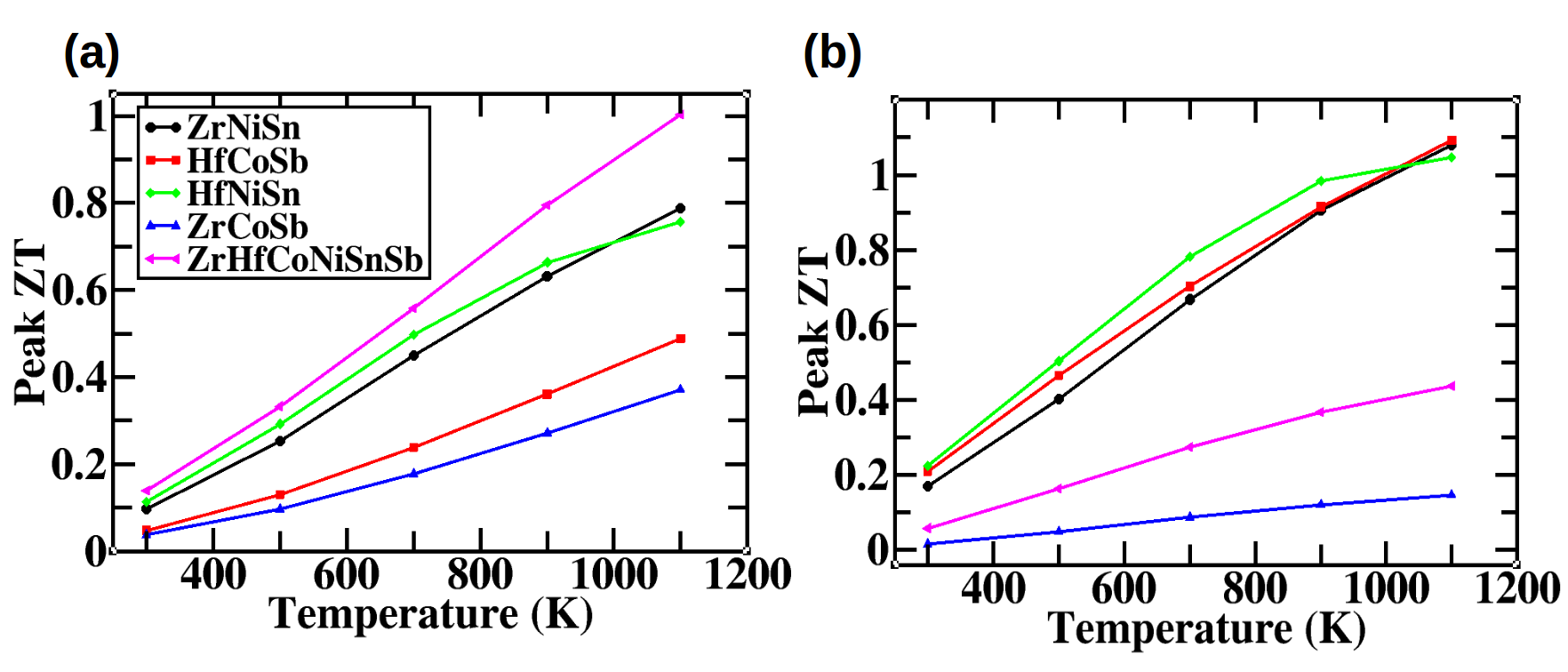} 
	 	 	\caption{(a) Peak value of figure of merit for electrons, (b) Peak value of figure of merit for  holes}
   \label{fig:peak_fom}
\end{figure}

\begin{figure}[h]
    \centering
\includegraphics[width=\columnwidth]{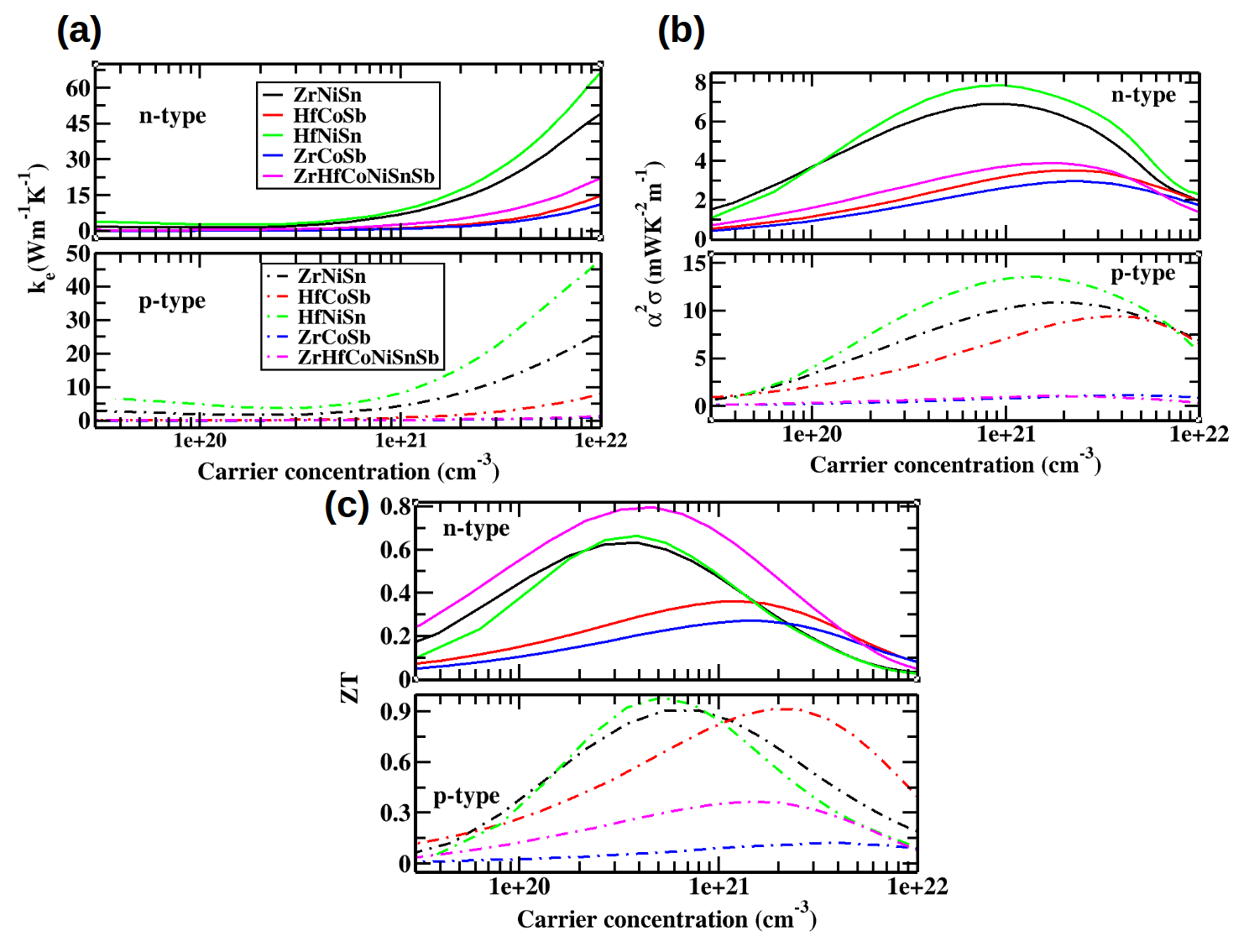} 
 	\caption{ (a) Electronic thermal conductivity, (b) Power factor and (c) Figure of merit as a function of carrier concentration at 900 K }
    \label{fig:other_trans_prop}
\end{figure}

The electronic transport properties, in combination with the lattice thermal conductivity, were used to compute the figure of merit (ZT) as a function of carrier concentration for different temperatures, as shown in Figure~\ref{fig:electronic_properties}(e). With increasing temperature, the ZT peak shifts toward higher carrier concentrations for both n-type and p-type carriers. At all temperatures, the ZT value for n-type carriers remains higher than that for p-type carriers. A comparison of ZT as a function of carrier concentration at 900 K among different structures is provided in the Supplementary Information (SI). As depicted in Figure~\ref{fig:other_trans_prop}(c), for n-type carriers, ZrHfCoNiSnSb exhibits the highest ZT value up to a carrier concentration of $4 \times 10^{21}$ cm$^{-3}$, followed by ZrNiSn and HfNiSn, which display nearly identical ZT values across all carrier concentrations. Despite their high Seebeck coefficients, HfCoSb and ZrCoSb exhibit relatively low ZT values for n-type carriers due to their lower electrical conductivity and higher lattice thermal conductivity, as seen in Figure~\ref{fig:other_trans_prop}(c) of SI. For p-type carriers, up to a concentration of $1 \times 10^{21}$ cm$^{-3}$, ZrNiSn and HfNiSn attain the highest and nearly identical ZT values, followed by HfCoSb. However, ZrHfCoNiSnSb and ZrCoSb display comparatively lower ZT values across all carrier concentrations, primarily due to their lower electrical conductivity. Figure~\ref{fig:peak_fom} presents a comparison of the peak ZT values for all structures at different temperatures. As observed, for n-type carriers, ZrHfCoNiSnSb consistently exhibits the highest peak ZT value across all temperatures, followed by HfNiSn and ZrNiSn. In contrast, HfCoSb and ZrCoSb show comparatively lower peak ZT values due to their lower electrical conductivity and higher lattice thermal conductivity. For p-type carriers, HfNiSn attains the highest peak ZT value, followed by HfCoSb and ZrNiSn, whereas ZrHfCoNiSnSb and ZrCoSb maintain lower peak ZT values across all temperatures due to their lower electrical conductivity. The peak ZT value of ZrHfCoNiSnSb for n-type carriers reaches 1.00 at 1100 K, which is 27, 104, 32, and 170 \% higher than that of ZrNiSn, HfCoSb, HfNiSn, and ZrCoSb, respectively. The optimized carrier concentration corresponding to this maximum ZT value at 300 K and 900 K is provided in Table~\ref{table:carr_con_p_type} and \ref{table:carr_con_p_type} in the Supplementary Information.

    \section{Conclusions}
      ZrHfCoNiSnSb exhibits greater stability at high temperatures compared to its parent compounds, largely due to the entropic contribution to the Gibbs free energy at elevated temperatures. The absence of imaginary modes in the phonon dispersion curve also confirms its dynamical stability. Additionally, based on both mode-resolved and average Gruneisen parameters, it is suggested that ZrHfCoNiSnSb possesses stronger anharmonicity and higher lattice thermal resistance than its parent compounds. The Seebeck coefficient and the other electronic transport properties of ZrHfCoNiSnSb exhibit comparability to the half-Heusler compounds from which it is composed(i.e., ZrNiSn/HfNiSn and HfCoSb/ZrCoSb). Notably, the lattice thermal conductivity of ZrHfCoNiSnSb is approximately one-third of  ZrNiSn/HfNiSn and one-fifth of HfCoSb/ZrCoSb at room temperature, and it has significantly reduced lattice thermal conductivity compared to the parent half-Heusler compounds across all temperatures. For the case when the charge carriers are electrons, at 1100 K, the ZT value of ZrHfCoNiSnSb is 1.00, surpassing the values of all of the parent compounds.
    
	 	 \clearpage

\bibliographystyle{aasjournal}
  \bibliography{references.bib}
\section{Supporting Information}
      \subsection{Crystal structures of the generated SQSs before geometric optimization.}

\clearpage

   \subsection{Phonon band structure, phonon DOS and IPR of the parent compounds}
\subsection{Band structure comparison of HEA for SOC and without SOC calculations}

\clearpage

\subsection{Orbital contributions of different elements}

\clearpage
\subsection{$\Delta \rho$ of the parent compounds}

\clearpage

\subsection{Effective masses of different bands}

The conduction band minimum (CBM) of all the parent compounds is located at the $\Gamma$ point and exhibits triple degeneracy. Among these three bands, two have light and equal effective masses, while the third has a heavy effective mass. In the case of the HEA, both the CBM and a conduction band extremum (CBE) are present, with the CBE positioned 37~meV above the CBM. The corresponding density-of-states and conductivity effective masses are listed in Tables~\ref{table:eff_mass_cbm_hh} and \ref{table:eff_mass_cbm_hea}.

\begin{table} [ht]
\centering 
\caption{Dos and conductive effective masses of conduction band minima (CBM) of the parent compounds}
\begin{tabular}{|c|c|c|c|c|}
\hline
Effective masses & ZrNiSn & HfNiSn & ZrCoSb & HfCoSb \\
\hline
$m_{D}^{*} $ (Heavy band) &3.33 &3.28 & 5.56 & 6.96 \\
\hline
 $m_{D}^{*} $ (two light band)  &0.42 &0.38 & 1.27 &1.09 \\
\hline
$m_{\sigma}^{*} $ (Heavy band) &3.33 &3.28 & 5.56 & 6.96 \\
\hline
 $m_{\sigma}^{*} $ (two light band)  &0.42 &0.38 & 1.27 &1.09 \\
\hline
\end{tabular}
\label{table:eff_mass_cbm_hh}
\end{table}

\begin{table} [ht]
\centering 
\caption{Dos and conductive effective masses of conduction band minima (CBM) of ZrHfCoNiSnSb}
\begin{tabular}{|c|c|c|}
\hline
Band & $m_{D}^{*}$ & $m_{\sigma}^{*}$ \\
\hline
CBM & 1.04 & 0.91 \\
\hline
CBE & 1.19 & 0.96 \\
\hline

\end{tabular}
\label{table:eff_mass_cbm_hea}
\end{table}

\begin{figure}[h]
    \centering
\includegraphics[width=\columnwidth]{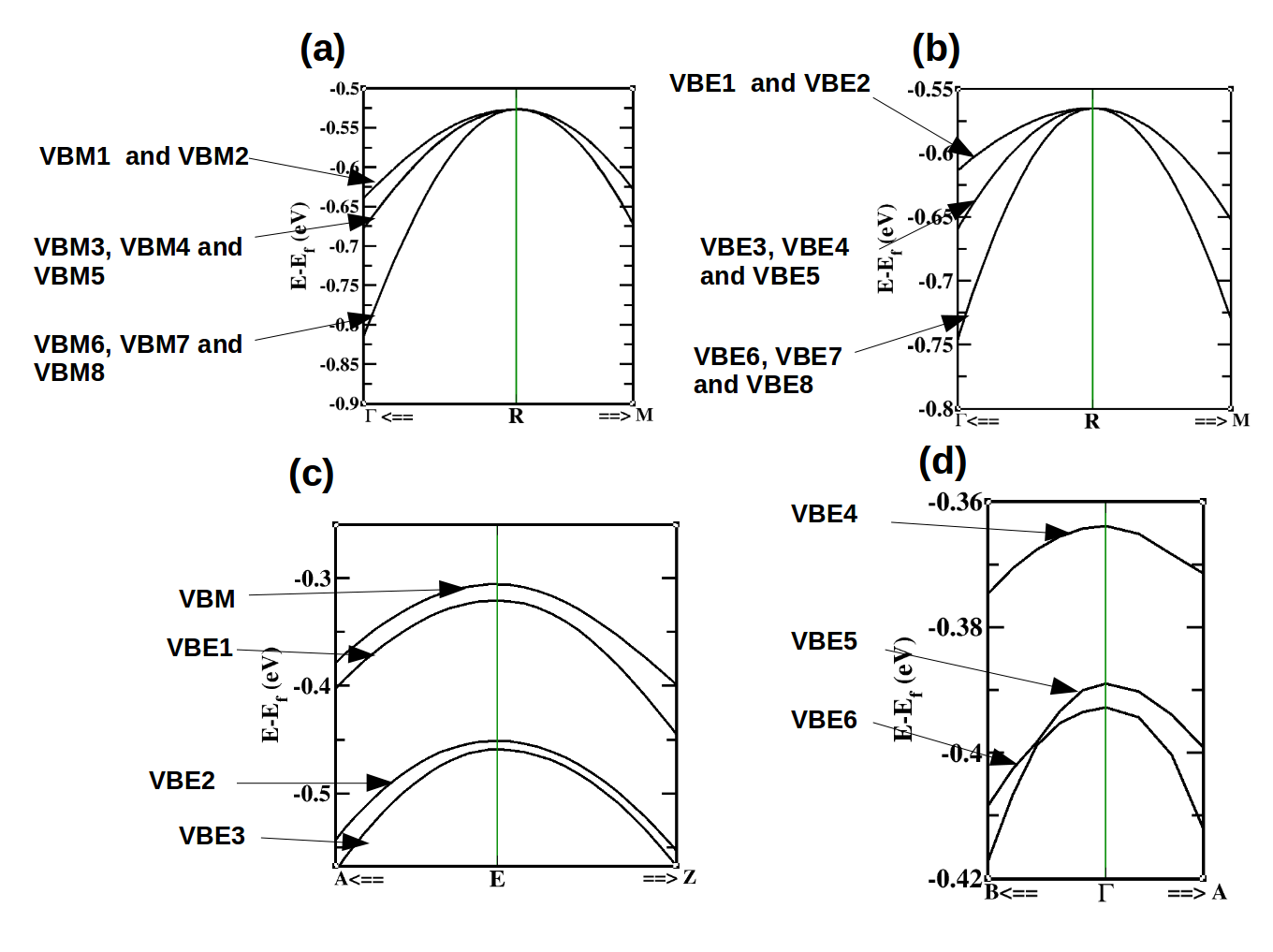} 
 	\caption{ (a) VBM of ZrCoSb at R point, (b) VBE of HfCoSb at $\Gamma$ point, (c) VBM/VBE of ZrHfCoNiSnSb at E point , and (d) VBE of ZrHfCoNiSnSb at $\Gamma$ point }
   \label{fig:effective_masses_plot}
\end{figure}

ZrNiSn and HfNiSn have their valence band maxima (VBM) at the $\Gamma$ point and do not exhibit a valence band extremum (VBE) elsewhere. Three bands contribute to the VBM in both cases: two heavy bands with equal effective masses and one light band. In HfCoSb also, the VBM is located at the $\Gamma$ point, with three bands contributing to it. Additionally, eight bands contribute to the VBE at the R point. For ZrCoSb, the VBM is located at the R point, with eight bands contributing to it. In the case of ZrHfCoNiSnSb, there are four bands at the E point, among which one corresponds to the VBM and the others to the VBE. Furthermore, three bands at the $\Gamma$ point are band extrema. The valley degeneracy at the R and E points is 8, which has been considered in calculating the DOS effective masses for the bands at these points.
For HfCoSb, the VBE lies 0.005 eV below the VBM. Band indices for different structures are shown in Figure~\ref{fig:effective_masses_plot}, and the corresponding effective masses are provided in Tables~\ref{table:eff_mass_vbm_zr_hf_nisn}, \ref{table:eff_mass_vbm_zrcosb}, \ref{table:eff_mass_vbm_hfcosb}, and \ref{table:eff_mass_vbm_hea}.

\begin{table} [ht]
\centering 
\caption{Dos and conductive effective masses of valence band maxima (VBM) of ZrNiSn and HfNiSn }
\begin{tabular}{|c|c|c|c|c|}
\hline
Effective masses & ZrNiSn & HfNiSn  \\
\hline
$m_{D}^{*} $ (Two heavy band) &0.87 & 0.69  \\
\hline
 $m_{D}^{*} $ (Light band)  &0.40 &0.37  \\
\hline
$m_{\sigma}^{*} $ (Two heavy band) &0.87 & 0.69  \\
\hline
 $m_{\sigma}^{*} $ (Light band)  &0.40 &0.37  \\
\hline
\end{tabular}
\label{table:eff_mass_vbm_zr_hf_nisn}
\end{table}

\begin{table} [ht]
\centering 
\caption{Dos and conductive effective masses of valence band maxima (VBM) of ZrCoSb}
\begin{tabular}{|c|c|c|}
\hline

Band & $m_{D}^{*} $ & $m_{\sigma}^{*} $ \\
\hline
VBM1 & 8.65 & 2.16 \\
\hline
VBM2 & 8.65 & 2.16 \\
\hline
VBM3 & 6.20 & 1.55 \\
\hline

VBM4 & 6.20 & 1.55 \\
\hline

VBM5 & 5.65 & 1.41 \\
\hline

VBM6 & 4.46 & 1.11 \\
\hline

VBM7 & 3.33 & 0.83 \\
\hline
VBM8 & 3.33 & 0.83 \\
\hline

\end{tabular}
\label{table:eff_mass_vbm_zrcosb}
\end{table}
\begin{table} [ht]
\centering 
\caption{Dos and conductive effective masses of VBM (at $\Gamma$ point) and VBE (at R point) of HfCoSb}
\begin{tabular}{|c|c|c|}
\hline

Band & $m_{D}^{*} $ & $m_{\sigma}^{*} $ \\
\hline
VBM(Two heavy bands) & 0.62 & 0.62 \\
\hline
VBM(One light bands) & 0.37 & 0.37 \\
\hline
VBE1 & 9.56 & 2.36 \\
\hline
VBE2 & 9.56 & 2.36 \\
\hline

VBE3 & 6.00 & 1.50 \\
\hline
VBE4 & 6.00 & 1.50 \\
\hline
VBE5 & 6.00 & 1.50 \\
\hline

VBE6 & 4.79 & 1.14 \\
\hline

VBE7 & 2.98 & 0.74 \\
\hline

VBE8 & 2.89 & 0.74 \\
\hline

\end{tabular}
\label{table:eff_mass_vbm_hfcosb}
\end{table}

\begin{table} [ht]
\centering 
\caption{Dos and conductive effective masses of valence band maxima (VBM) at E point and VBE at E point and $\Gamma$  point of ZrHfCoNiSnSb}
\begin{tabular}{|c|c|c|c|}
\hline

Band & $m_{D}^{*} $ & $m_{\sigma}^{*} $ & $\Delta E$\\
\hline

 VBM & 5.40 & 1.34 & 0 \\
 \hline
 
 VBE1 & 4.62 & 1.15 & 0.015 \\
 \hline
 
 VBE2 & 5.64 & 1.33 & 0.145 \\
 \hline

  VBE3 & 4.23 & 1.06 & 0.153 \\
 \hline
 VBE4 & 0.66 & 0.62 & 0.059 \\
 \hline
 
  VBE5 & 0.69 & 0.61 & 0.084 \\
 \hline
 VBE6 & 0.52 & 0.50 & 0.088 \\
 \hline
\end{tabular}
\label{table:eff_mass_vbm_hea}
\end{table}

\clearpage

   \subsection{Transport properties comparison at 900 K}
     \begin{figure}[h]
    \centering
\includegraphics[width=\columnwidth]{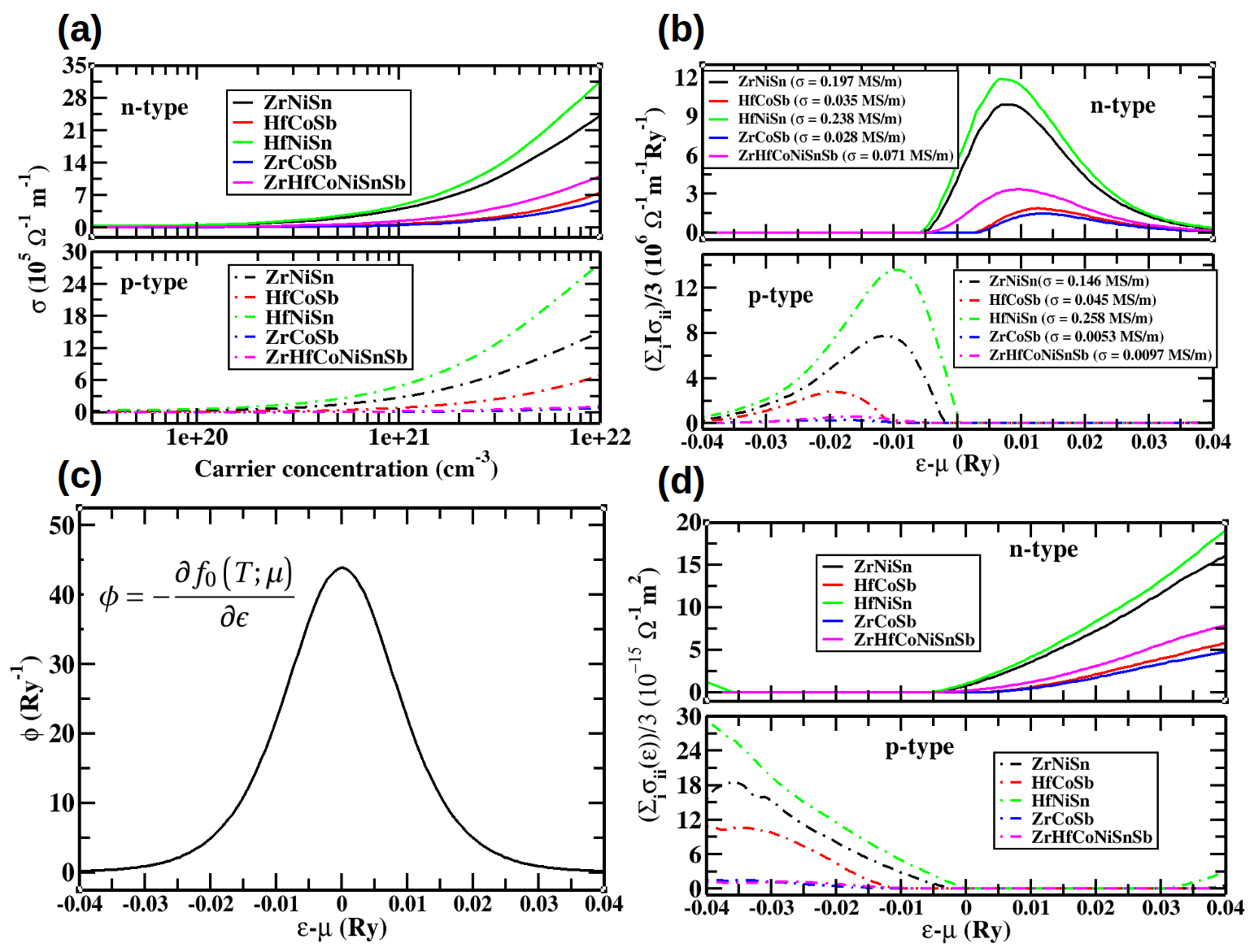} 
 	\caption{ (a) Electrical conductivity as a function of carrier concentration, (b)  average of integrand of electrical conductivity expression as a  function of $\epsilon - \mu$, (c) selection function $\phi$ as a function of $\epsilon - \mu$, and (d) Average of projected conductivity tensor as a function of $\epsilon-\mu$ }
  \label{fig:ep_comp_ec}
\end{figure}

Figure~\ref{fig:ep_comp_ec} (a) presents the variation in electrical conductivity with carrier concentration at 900 K for both the parent compounds and HEA. In the case of n-type conductivity, HfNiSn exhibits the highest electrical conductivity across all carrier concentrations, followed by ZrNiSn, ZrHfCoNiSnSb, HfCoSb, and ZrCoSb in descending order. For the p-type case, a similar trend is observed, with HfNiSn showing the highest conductivity, followed by ZrNiSn, HfCoSb, ZrHfCoNiSnSb, and ZrCoSb.  

This trend in electrical conductivity can be understood by analyzing the integrand of the expression of electrical conductivity $ \sigma_{ij} \left(T;\mu \right) $ (refer to equation number) for different values of the chemical potential $ \mu $. Figure~\ref{fig:ep_comp_ec}(b) illustrates the integrand of the averaged electrical conductivity tensor I$ \sigma_{ii} \left(T;\mu \right) $ as a function of $ \epsilon - \mu $, where $ \mu $ corresponds to a carrier concentration of $ 5\times 10^{20} $ cm$^{-3}$. The total electrical conductivity is determined by the integral of these curves, representing the area under them.  This specific carrier concentration is chosen because the figure of merit for ZrHfCoNiSnSb reaches its maximum around this value for the n-type case at 900K, making it an interesting point for investigating transport properties. Similar analyses can be conducted for other carrier concentrations.  For the n-type case, the chemical potential for this carrier concentration is 81 meV above CBM for ZrNiSn, 90 meV above CBM for HfNiSn, and 80 meV above CBM for ZrHfCoNiSnSb, while for HfCoSb and ZrCoSb, it lies 30 meV and 35 meV below CBM, respectively, within the bandgap. Similarly, for the p-type case, it is 3 meV below VBM for HfNiSn, whereas for ZrNiSn, HfCoSb, ZrCoSb, and ZrHfCoNiSnSb, it is positioned 25 meV, 140 meV, 115 meV, and 40 meV above VBM, respectively, within the bandgap.

The electrical conductivity integrand, I$ \sigma_{ii}\left(T;\mu\right) $, consists of the projected conductivity $ \sigma_{ii}(\epsilon) $, shown in Figure~\ref{fig:ep_comp_ec} (d), weighted by the selection function $ \phi $, which is plotted in Figure~\ref{fig:ep_comp_ec}(c). The projected conductivity at different energy values is calculated using the BoltzTraP1 code. Since the selection function decreases to approximately $99\%$ of its peak value around $\pm0.04$ Ry from the chemical potential, the most significant contribution of the projected conductivity tensor to electrical conductivity arises within this energy range. 

As depicted in Figure~\ref{fig:ep_comp_ec}(d), for the n-type case, the projected conductivity within this energy window is highest for HfNiSn, followed by ZrNiSn, ZrHfCoNiSnSb, HfCoSb, and ZrCoSb. For the p-type case, HfNiSn again exhibits the highest projected conductivity, followed by ZrNiSn, HfCoSb, ZrHfCoNiSnSb, and ZrCoSb. Consequently, across the entire relevant energy range around the chemical potential, the trend in projected conductivity aligns with the trend observed in electrical conductivity (Figure~\ref{fig:ep_comp_ec}(a)). Thus, for both the n-type and p-type cases, the order of electrical conductivity follows from the corresponding projected conductivity, reinforcing the observed hierarchy among the compounds.

     \begin{figure}[h]
    \centering
\includegraphics[width=\columnwidth]{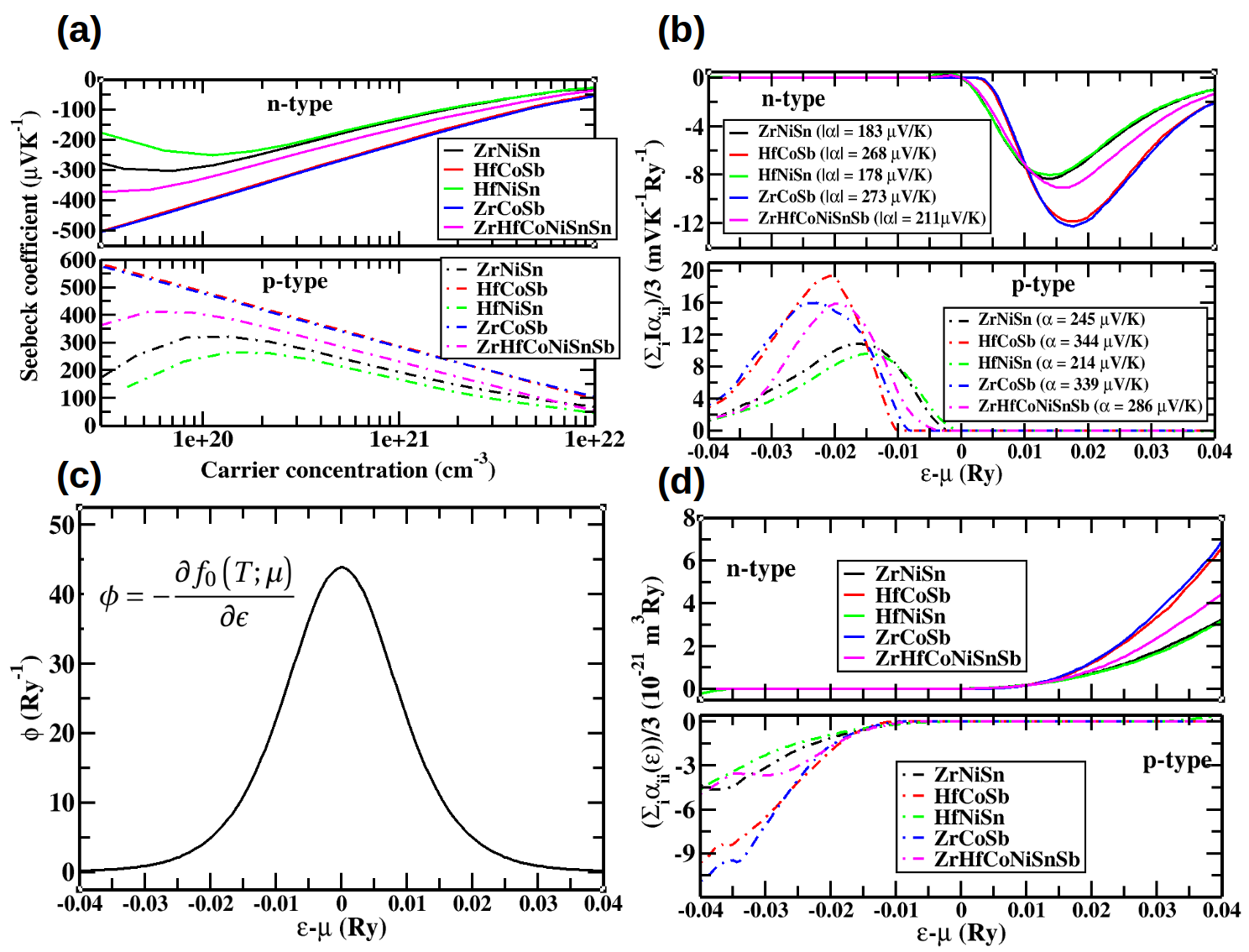} 
 	\caption{(a) Seebeck coefficient as a function of carrier concentration, (b)  average of integrand of Seebeck coefficient expression as a  function of $\epsilon - \mu$, (c) selection function $\phi$ as a function of $\epsilon - \mu$, and (d) Average of projected Seebeck tensor as a function of $\epsilon-\mu$  }
   \label{fig:ep_comp_seebeck}
\end{figure}

Figure~\ref{fig:ep_comp_seebeck} (a) presents the Seebeck coefficient as a function of carrier concentration at 900 K for different structures. From the plot, it is evident that for the n-type case, ZrCoSb exhibits the highest Seebeck coefficient, followed by HfCoSb, ZrHfCoNiSnSb, ZrNiSn, and finally HfNiSn. For the p-type case, HfCoSb has the highest Seebeck coefficient, followed by ZrCoSb, ZrHfCoNiSnSb, ZrNiSn, and HfNiSn.  

Similar to the previous case, the integrand of the expression for the Seebeck coefficient, I\( \alpha_{ii}\left(T;\mu\right) \) (refer to equation number), is plotted in  Figure~\ref{fig:ep_comp_seebeck}(b) as a function of \( \epsilon - \mu \) for the chemical potential corresponding to a carrier concentration of \( 5 \times 10^{20} \) cm\(^{-3}\). The Seebeck coefficient is determined by the area under these curves. This integrand primarily consists of the projected Seebeck tensor \( \alpha_{ii}(\epsilon) \), weighted by the selection function. The projected Seebeck  tensor is defined as:

\begin{equation}
    \alpha_{ii}(\epsilon) = \sum_{k} \sigma_{ik}^{-1} \left(T;\mu\right) \left(\epsilon - \mu \right) \sigma_{ki} (\epsilon)
        \label{eq:projected_seebeck_tensor}
\end{equation}

Figure~\ref{fig:ep_comp_seebeck} (d) illustrates the plot of the average projected Seebeck tensor as a function of \( \epsilon - \mu \) for different structures. As observed from the plot, for the n-type case, the average projected Seebeck tensor attains its maximum value within the relevant energy range (\(-0.04\) to \(0.04\) Ry around \( \mu \)) for ZrCoSb, followed by HfCoSb, ZrHfCoNiSnSb, ZrNiSn, and HfNiSn. This trend in the projected Seebeck tensor directly explains the ordering of the Seebeck coefficient for the n-type case, as shown in Figure~\ref{fig:ep_comp_seebeck}(a).  

For the p-type case, although there are fluctuations in the projected Seebeck tensor values between ZrCoSb and HfCoSb, as well as between ZrHfCoNiSnSb and ZrNiSn, the dominant energy region where the selection function has significant weight (\(-0.02\) to \(0\) Ry around \( \mu \)) shows HfCoSb having the highest projected Seebeck tensor value, followed closely by ZrCoSb. This indicates that the Seebeck coefficient of HfCoSb is slightly higher than that of ZrCoSb.  

Although ZrHfCoNiSnSb has a projected Seebeck tensor value close to that of ZrCoSb in the main energy range, for energies lower than \(-0.02\) Ry about \( \mu \), the projected Seebeck tensor for ZrCoSb exceeds that of ZrHfCoNiSnSb. This explains why ZrCoSb has a higher Seebeck coefficient than ZrHfCoNiSnSb in the p-type case. Similarly, within the \(-0.02\) to \(0\) Ry energy range around \( \mu \), the projected Seebeck tensor for ZrHfCoNiSnSb is greater than that of ZrNiSn, followed by HfNiSn, explaining the observed trend among these structures.  \\

\clearpage
\subsection{Carrier concentration for peak value of figure of merit}
\begin{table} [ht]
\centering 
\caption{Carrier concentration for n-type case for the peak value of the figure of merit (in the unit of 10$^{20}$ cm$^{-3}$) }
\begin{tabular}{|c|c|c|c|c|c|}
\hline
Temperature & ZrNiSn & HfCoSb  & HfNiSn & ZrCoSb & ZrHfCoNiSnSb \\
\hline
300 K & 1.16 & 5.68 & 1.14 & 5.82 & 1.68 \\
\hline
900 K & 3.87 & 13.4 & 3.88 & 12.8 & 4.66 \\
\hline

\end{tabular}
\label{table:carr_con_n_type}
\end{table}

\begin{table} [ht]
\centering 
\caption{Carrier concentration for the p-type case for the peak value of figure of merit (in the unit of 10$^{20}$ cm$^{-3}$) }
\begin{tabular}{|c|c|c|c|c|c|}
\hline
Temperature & ZrNiSn & HfCoSb  & HfNiSn & ZrCoSb & ZrHfCoNiSnSb \\
\hline
300 K & 2.02  & 10.5 & 1.60 & 57.2 & 11.1 \\
\hline
900 K & 7.92 & 18.2 & 5.14  & 43.1 & 14.1 \\
\hline

\end{tabular}
\label{table:carr_con_p_type}
\end{table}

\section{Acknowledgment}

The author gratefully acknowledges Dr. Prasenjit Ghosh for his valuable guidance in structuring the results and the manuscript, as well as for providing access to computational resources. The author also thanks the National Supercomputing Mission (NSM) for providing computing facilities on the "Param Brahma" supercomputer at IISER Pune. The author further acknowledges Dr. Gautam Sharma from Khalifa University for many insightful and fruitful discussions and IISER Pune for fellowship.

\clearpage


\end{document}